\documentclass{article}
\usepackage{diagrams}
\usepackage{latexsym,amssymb, amsfonts, amscd}

\input diagrams

\def\Sp{\hbox{Sp}}
\def\Bbb{\mathbb}
\newtheorem{proposition}{Proposition}
\newtheorem{corollary}{Corollary}
\newtheorem{lemma}{Lemma}
\newtheorem{remark}{Remark}
\newtheorem{definition}{Definition}
\newtheorem{theorem}{Theorem}

\begin{document}
\title{Holonomy and Skyrme's model}
%\subtitle{Do you have a subtitle?\\ If so, write it here}
\author{Dave Auckly \thanks{{The first author was partially supported by 
NSF grant DMS-0204651.}} 
\and Lev Kapitanski% etc
% \thanks is optional - remove next line if not needed
\thanks{{  
The second author was partially supported by NSF grants 
DMS-9970638, and DMS-0200670} }%
}                     % Do not remove
\date{Department of Mathematics, \\ Kansas State University, \\ 
Manhattan, Kansas 66506, USA}
%
%\date{Received: date / Accepted: date}
% 
%
%
\maketitle
\begin{abstract} 
In this paper we consider two generalizations of the Skyrme model. 
One is a variational problem for maps from a compact $3$-manifold 
to a compact Lie group. The other is a variational problem 
for flat connections. 
We describe the path components of the configuration 
spaces of smooth fields for each of the variational problems. 
We prove that the invariants separating the path components are 
well-defined for (not necessarily smooth) fields with finite Skyrme energy. 
We prove that for every possible value of these invariants 
there exists a minimizer of the Skyrme functional. 
Throughout the paper we emphasize the importance of 
holonomy in the Skyrme model. Some of the results may be useful in 
other contexts. In particular, we define the holonomy of 
a distributionally flat $\,L^2_{loc}\,$ connection; the local 
developing maps for such connections need not be continuous. 
\end{abstract}
%
%%%%%%%%%%%%%%%%%%%%%%%%%%%%  INTRO %%%%%%%%%%%%%%%%
\section{Introduction}
\label{intro}
In 1961 T. H. R. Skyrme introduced a model to describe self-interacting 
meson fields, \cite{Skyrme1}, \cite{Skyrme2}, \cite{Skyrme3}. 
For a review of the physical and mathematical literature on the Skyrme model 
see \cite{ZB}, \cite{Gisiger},  \cite{LK}.

The static fields of the original Skyrme model may be described as  
maps $\,u\,$ from $\,\mathbb R^3\,$ into the group of unit quaternions, 
$\,\Sp(1)\simeq SU(2)$. It is required that these maps satisfy 
the boundary condition  
$\,u(\infty)= 1\,$ and have finite energy
$$
E(u)=\int_{\Bbb R^3} \frac12 |u^{-1}\,d u|^2\,
+\,\frac14 |u^{-1}\,d u\wedge u^{-1}\,d u|^2\;dx\,,
$$
where 
$$
|u^{-1}\,d u|^2\,=\,\sum_{j=1}^3 
|u^{-1}\,{\partial u\over\partial x^j}|^2\;,
$$
$$
|u^{-1}\,d u\wedge u^{-1}\,d u|^2\,=\,
\sum_{j<k}
|\left[ u^{-1}\,{\partial u\over\partial x^j},
\,u^{-1}\,{\partial u\over\partial x^k}\right]|^2\,,
$$
and $\,|q|^2=q\cdot\bar q = \sum_{j=1}^4(q_j)^2$. In addition, 
the space of smooth maps satisfying the boundary condition 
splits into infinitely many different components 
classified by the degree of the map. 

In the very first paper \cite{Skyrme1}, p.129, Skyrme noted that 
$\,u^{-1}du\,$ is a flat connection.  In terms of the connection 
$\,a = u^{-1} du\,$ the energy takes the form 
$$
E[a]=\int_{\Bbb R^3} \frac12 |a|^2\,+\,\frac1{16} |[a,\,a]|^2\;dx\,.
$$
Some generalizations of this functional have been considered 
previously, \cite{Manton}, \cite{Loss}. 
It is natural to generalize 
the original setting in several directions. First, one may replace 
$\,\mathbb R^3\,$ with an arbitrary Riemannian 3-manifold. Second, 
one may consider maps into arbitrary Lie groups. Third, the model may 
be described in terms of flat connections on principal $G$-bundles.

If the manifold, $\,M^3\,$ is non-compact, the existence of ground states
is a difficult open problem. On the other hand, if $\,M^3\,$ is compact 
(and $\,G=SU(2)$), 
the existence of ground states is much easier and has been 
established in \cite{LK}. The purpose of the present paper is to understand 
the underlying geometry of the space of maps and/or the space of connections 
together with the interaction between this geometry and the analytical 
behavior of the Skyrme fields. All the geometric features are present 
for closed 3-manifolds. For this reason we restrict to the case of 
closed 3-manifolds. The density 
$\,\frac12 |u^{-1}\,d u|^2\,
+\,\frac14 |u^{-1}\,d u\wedge u^{-1}\,d u|^2\,$ requires a metric 
on the Lie group. For compact Lie groups the Killing metric 
is bi-invariant. 
This leads to a symmetry of the Skyrme functional. In addition, maps into
a compact Lie group are automatically bounded.
For these reasons we 
restrict our attention to compact Lie groups. 

In order to retain similarity with the original model, any generalization 
should be invariant under the group. For maps this means that 
$\,E(u\cdot g)=E(u)\,$ for any constant $\,g\in G$. For connections, 
this means that the constant gauge transformations should contain a copy 
of the group $\,G$ and the energy should be invariant under the constant 
gauge transformations.

A connection on a principle $G$-bundle, $\,P\to M\,$ is a Lie algebra 
valued 1-form, $A$, on the total space of the bundle, \cite{KN}. 
The energy 
density $\,\frac12 |A|^2\,+\,\frac1{16} |[A,\,A]|^2\,$ is thus a function 
on the total space, $\,P$. This function is not $G$-equivariant, so it does 
not descend to a function on $M$. One way to resolve this problem is to pick 
 a reference connection, $B$, on $\,P$, and write $A=B+a$, where $a$ 
may uniquely be identified as an element of 
$\,\Gamma(T^*M\otimes\hbox{Ad}_*P)$, \cite{DK}. Thus, 
$\,\mathcal L(a)=\frac12 |a|^2\,+\,\frac1{16} |[a,\,a]|^2\,$ 
is a well-defined density on $M$. 
The gauge group is the group of automorphisms of the bundle. It may be 
identified with $\,\Gamma(\Lambda^0M\otimes\hbox{Ad}\,P)$. 
Then a gauge transformation, $\,g$, acts on a twisted 1-form, $\,a$, via 
$\,a\cdot g=g^{-1} a g + g^{-1} d_B g$. Note that the density 
$\,\mathcal L(a)$ is {\it not} gauge invariant unless the gauge transformation 
is $\,B$-covariantly constant ($\,d_B g = 0$). Indeed, 
$\,\mathcal L(0\cdot g)=\mathcal L(0)=0\,$ implies 
$\,
\frac12 |d_B g|^2\,+\,\frac1{16} |[d_B g,\,d_B g]|^2 =0$.
The energy density {\it is} invariant under $\,B$-covariantly 
constant gauge transformations. The $\,B$-covariantly 
constant gauge transformations can be identified with the cenralizer 
of the holonomy of $\,B$, \cite{DK}. Thus, the energy density is 
invariant under the group exactly when $B$ has central holonomy. 
By the holonomy reduction theorem, \cite{KN}, there is a subbundle $Q$ of $P$ 
with structure group $\,Z$, the center of $G$, so that 
$\,\hbox{Ad}_*P\cong Q\times_Z \mathfrak g$. 
Since $Z$ acts trivially on the Lie algebra $\mathfrak g$, 
we have an isomorphism $\,Q\times_Z \mathfrak g\cong M\times \mathfrak g$, 
i.e., $\,[(q, X)]\mapsto ([q], X)$. In other words, 
the gauge group becomes $\,\hbox{Maps}(M,G)\,$ in our case.
The flat connections on $P$ correspond 
to the set 
$\,\{a\in\Gamma(T^*M\otimes\mathfrak g)\,|\;da+\frac12 [a,a]+F_B=0\}$, 
where $\,F_B$ is the curvature of the reference connection $B$ 
regarded as an element of $\,\Gamma(\Lambda^2M\otimes\mathfrak g)$. 
Given a unitary representation $\,\alpha:\,G\to U(n)$, the associated 
complex vector bundle $\,P\times_\alpha \mathbb C^n\,$ has 
first Chern class $\,-\frac{1}{2\pi i}\hbox{Trace}(\alpha(F_B))$. 
It follows that the 
curvature of $B$ is a closed central 2-form with integral periods. 
Conversely, any closed central 2-form with integral periods arises 
as the curvature of the cenral connection on the bundle. 

To summarize, we have two distinct variational problems, 
one for equivalence classes of 
maps from a closed 3-manifold to a compact Lie group, 
and one for flat connections $\,A=B+a$ modulo $B$-covarianly constant 
gauge transformations. 

In this paper we describe the path components of the configuration 
spaces of smooth fields for each of the above variational problems. 
We prove that the invariants separating the path components are 
well-defined for (not necessarily smooth) fields with finite Skyrme energy. 
We prove that for every possible value of these invariants 
there exists a minimizer of the Skyrme functional. 
Throughout the paper we emphasize the importance of 
holonomy in the Skyrme model. Some of the results may be useful in 
other contexts. In particular, we define the holonomy of 
a distributionally flat $\,L^2_{loc}\,$ connection; the local 
developing maps for such connections need not be continuous. 

We now describe the contents of the paper in more detail. 
In Section 2 below we prove that the homotopy classes of maps from 
$M$ to $G$, $\,[M,\,G]$, are in bijective correspondence with 
$\,G/G_0\times H^3(M, H_3(\tilde G))\times H^1(M,H_1(G_0))$. 
Here $\,G_0\,$ is the identity component of $G$, and 
$\,\tilde G\,$ is the universal 
covering group of $\,G_0$. Since the energy, $E(u)$, is $G$-invariant, 
the relevant set of classes of maps is $\,[M,\,G]/G$. This set 
is isomorphic to 
$\,H^3(M, H_3(\tilde G))\times H^1(M,H_1(G_0))$. We also give 
a concrete analytic description of the corresponding invariants in Sections 
2, 3 and 4.  

The topological type of a connection modulo $B$-covarianly 
constant gauge
is specified by a Chern-Simons invariant and a holonomy representation. 
These correspond to 
$\,H^3(M, H_3(\tilde G))\,$ and $\, H^1(M,H_1(G_0))$, respectively. 
 The Chern-Simons invariant 
is well defined for connections with finite Skyrme energy. 
In Section 3 we give a definition of holonomy 
that is valid for distributionally flat 
$\,L^2_{loc}\,$ connections. In particular, 
our definition is valid  for flat connections with bounded Skyrme energy.

This generalization of holonomy requires a 
nonlinear version of Poincar\'e's lemma. More precisely, 
in Section 3 we prove that any distributionally flat $L^2_{loc}$ connection 
is locally trivial. For more regular connections (namely, $\,A\in W^{1,3/2}$), 
this follows from a theorem of K. Uhlenbeck, 
\cite{U1}. 

In Section 4 we prove that the invariants are well defined for maps with bounded 
Skyrme energy. 
In Section 5 we consider the minimization problems for the 
Skyrme functionals. In particular, we prove that for any fixed set of 
invariants there exists a map attaining these values 
that minimizes the Skyrme energy $\,E(u)\,$ in this class. 
Moreover, we also solve the analogous problem in the space of flat connections.

%%%%%%%%%%%%%%%%%%%%%%%%%%%%% SECTION 2 %%%%%%%%%%%%%%%%%%%%%

\section{Homotopy classes of maps}\label{sec:2}

Let $\,M\,$ be a closed 3-manifold.
Let $G$ be a comact Lie group, $\,G_0\,$ its identity component, and  
$\,\tilde G\,$ the universal covering group of $G_0$, with covering 
projection $\,p:\,\tilde G\to G_0$. 
Denote by $\,[M,\,G]\,$ the free homotopy classes of continuous 
maps from $\,M\,$ to $\,G$. 

\subsection{Algebraic description}
\label{sec:2.1}
\begin{proposition}\label{prop1} As sets, 
$$
[M,\,G]\,\cong\,G/G_0\,\times H^3(M; H_3(\tilde G))\times H^1(M; H_1(G_0))\,.
$$
\end{proposition}
\noindent{\bf Proof.} Pick a point $x_0\in M$. There is an isomorphism 
$\,\phi_0:\,[M,\,G]\to G/G_0\times [M,\,G_0]\,$ given by 
$\,\phi_0([u])=([u(x_0)],[u(x_0)^{-1}\cdot u])$. Now we will regard 
$\,[M,\,G_0]\,$ as based homotopy classes. 

The homotopy classes of maps into a Lie group form a group 
under pointwise multiplication. The covering projection, 
$\,p:\,\tilde G\to G_0$, induces a homomorphism 
$\,p_*:\,[M,\,\tilde G]\to [M,\,G_0]$ by composition. 
The homomorphisms  from $\,\pi_1(M)\,$ to $\,\pi_1(G_0)\,$ 
form a group under pointwise multiplication. 
The natural map,  
$\,\pi_1:\;[M,\,G_0]\to \hbox{Hom}(\pi_1(M, x_0),\,\pi_1(G_0, 1))$,  
is a group homomorphism. Consider the sequence
\begin{equation}\label{seq1}
1\to [M,\,\tilde G]\,\stackrel{ p_*}{\rightarrow}\,[M,\,G_0]\,
\stackrel{ \pi_1}{\rightarrow}\,
\hbox{Hom}(\pi_1(M),\,\pi_1(G_0))\to 1 \,.
\end{equation}
If $\,p_*(\bar u) = 1$, there exists a homotopy 
$\,H:\,M\times I\to G_0\,$ to the constant map making the following diagram 
commute:
\newarrow{Dashto}....>
\begin{diagram}
M & \rTo^{\bar u} & \tilde G \\
\dTo^{i_0} & \ruDashto^{\bar H} & \dTo_{p} \\
M\times I  & \rTo_{H} & G_0\\
\end{diagram} 
By the homotopy lifting theorem, \cite{Spanier}, there is 
a homotopy $\,\bar H\,$ that makes the extended diagram commute.  
By the unique lifting theorem, \cite{Spanier}, $\,\bar H_1\,$ is constant; 
hence, $\,p_*\,$ is injective. 

Since $\,\tilde G\,$ is simply connected, $\,\pi_1(p_*\circ \bar u)\,$ 
is the trivial homomorphism for all $\,\bar u\in [M,\,\tilde G]$. So, 
the image of $\,p_*\,$ lies in the kernel of $\,\pi_1$. 
If $\,\pi_1(u)=1$, there is a map $\,\bar u\,$ so that 
$\,p_*\circ \bar u = u$, by the lifting theorem, \cite{Spanier}. 
Thus, the sequence (\ref{seq1}) is exact at $\,[M,\,\, G_0]$.

Any closed, connected 3-manifold admits a Heegaard splitting, 
 \cite{Rolfsen}, \cite{Moise}, 
and, therefore, a CW decomposition with exactly one 0-cell and one 3-cell. 
Given a homomorphism, $\,\alpha:\,\pi_1(M, x_0)\to \pi_1(G_0,1)$, 
construct a map $\,u_\alpha: M\to G_0\,$ as follows. On the 0-skeleton, 
define $\,u_\alpha^{(0)}(x_0)=1$. Any 1-cell of $\,M\,$ specifies 
an element of $\,\pi_1(M)$. Pick a representative of the corresponding class 
in $\,\pi_1(G_0)\,$ and define $\,u_\alpha^{(1)}\,$ on the 1-cell 
via this representative. The attaching map of any 2-cell is trivial in 
$\,\pi_1(M)$, so the composition with $\,u_\alpha^{(1)}\,$ is trivial in 
$\,\pi_1(G_0)$ and thus extends to a map of the disk into $\,G_0$. 
Define $\,u_\alpha^{(2)}\,$ on the 2-cell via this map. 
The composition of the attaching map of the 3-cell with 
$\,u_\alpha^{(2)}\,$ is trivial in $\,\pi_2(G_0)\,$ since $\,\pi_2\,$ 
of any Lie group is trivial, \cite{Br}. Thus, the composition extends 
to a map of the 3-disk that may be used to define the map 
$\,u_\alpha=u_\alpha^{(3)}$. By construction, $\,\pi_1([u_\alpha])=\alpha$, 
i.e., $\,\pi_1\,$ is surjective. Thus, the sequence (\ref{seq1}) 
is exact. 

Given any short exact sequence of groups 
$\,1\to K \to G \to H \to 1$, there is a bijection $\,G\cong K\times H$. 
Thus, 
$\,[M,\,G_0]\cong [M,\,\tilde G]
\times \hbox{Hom}(\pi_1(M, x_0),\,\pi_1(G_0, 1))$. 

Let $\,M^{(k)}\,$ denote the $k$-skeleton of $M$ with base point $x_0$. 
Let $\,SX\,$ 
denote the suspension of $(X, x_*)$, i.e., the cylinder $\,X\times [0,\,1]\,$ 
with both ends and the segment $\,\{x_*\}\times [0,\,1]\,$ 
collapsed to a single point. 
 
Consider the sequence
\begin{equation}\label{seq2}
\begin{diagram}
M^{(2)} & \rTo^{i_2} & M^{(3)} & \rTo^{q_2} & M^{(3)}/M^{(2)} & 
\rTo^{\partial_2} & SM^{(2)} & \rTo & SM^{(3)}. \\
\end{diagram}
\end{equation}
The space $\,M^{(3)}/M^{(2)}\,$ is homeomorphic to $\,D^3/S^2$. Under this 
identification, 
$\,\partial_2(x) = \left(f^{(3)}(\frac{x}{|x|}), |x|\right)$ for $\,x\in D^3$, 
where $\,f^{(3)}:\,S^2\to M^{(2)}\,$ is the attaching map for the 3-cell. 
The sequence (\ref{seq2}) is co-exact, \cite{Spanier}. 
Therefore, it induces the exact sequence 
$$
[SM^{(2)},\,\tilde G]\to [M^{(3)}/M^{(2)},\,\tilde G]\to [M,\,\tilde G]\,
\to [M^{(2)},\,\tilde G] = 0.
$$
Similarly, the sequence 
\begin{equation} \label{seq3}
\begin{diagram}  
M^{(1)} & \rTo & M^{(2)} & \rTo & M^{(2)}/M^{(1)} & 
\rTo & SM^{(1)} & \rTo & \\ 
\rTo & SM^{(2)} & \rTo & 
S(M^{(2)}/M^{(1)}) & 
\rTo & S^2M^{(1)} \\
\end{diagram}
\end{equation}
induces the exact sequence 
$$
[S^2M^{(1)},\,\tilde G]\to [S(M^{(2)}/M^{(1)}),\,
\tilde G]\to [SM^{(2)},\,\tilde G]\,
\to [SM^{(1)},\,\tilde G] = 0.
$$
These exact sequences may be spliced together to give 
\begin{diagram}\label{}
\qquad[SM^{(2)},\,\tilde G]\quad & \rTo^{\partial^*_2} & 
[M^{(3)}/M^{(2)},\,\tilde G] & \rTo^{q^*_2} & [M,\,\tilde G]
& \rTo & 0 \\ 
 \uTo_{q^*_1} & \ruTo & & & & &  \\ 
[S(M^{(2)}/M^{(1)}),\,\tilde G] & & & & & & \\ 
\end{diagram}
Thus,
\begin{eqnarray*} 
[M,\,\tilde G]& \cong [M^{(3)}/M^{(2)},\,\tilde G]/ \hbox{Ker}\, q^*_2 
\cong [M^{(3)}/M^{(2)},\,\tilde G]/ \hbox{Im}\, \partial^*_2 \\
& \cong [M^{(3)}/M^{(2)},\,\tilde G]/ \hbox{Im}\, (\partial^*_2\circ q^*_1)\,.
\end{eqnarray*}
The sequences (\ref{seq2}), (\ref{seq3}) induce 
$$
H_3(M^{(3)})\stackrel{q_{2 *}}{\rightarrow} 
H_3(M^{(3)}/M^{(2)})\stackrel{\partial^3_{*}}{\rightarrow} H_2(M^{(2)})
$$
and 
$$
H_2(M^{(2)})\stackrel{q_{1 *}}{\rightarrow} 
H_2(M^{(2)}/M^{(1)})\stackrel{\partial^2_{*}}{\rightarrow} H_1(M^{(1)})\,, 
$$ 
which give 
$$
H_3(M^{(3)}/M^{(2)})\stackrel{q_{1 *}\circ\partial^3_{*}}{\rightarrow} 
H_2(M^{(2)}/M^{(1)})
$$ 
and the commutative diagram 
\begin{equation} \label{diag2}
\begin{diagram}
\hbox{Hom}(H_2(M^{(2)}/M^{(1)}),\pi_3(\tilde G)) & 
\rTo^{q_{1 *}\circ\partial^3_{*}} & 
\hbox{Hom}(H_3(M^{(3)}/M^{(2)}),\pi_3(\tilde G)) \\ 
\uTo^{F_2} & & \uTo_{F_3} \\ 
[S(M^{(2)}/M^{(1)}),\tilde G] & \rTo_{\partial^*\circ q_1^*} & 
[M^{(3)}/M^{(2)},\tilde G]\\
\end{diagram}
\end{equation}
Here $\,F_3([u])[\gamma] = [u\circ r_3^{-1}(\gamma)]\,$ and 
$\,F_2([u])(\gamma) = [u\circ r_2^{-1}\circ (\partial^s_*)^{-1}(\gamma)]$, 
where $\,\partial^s_*:\,H_3(S(M^{(2)}/M^{(1)}))\to H_2(M^{(2)}/M^{(1)})\,$ 
is the suspension isomorphism, and 
$\,r_3:\,\pi_3(M^{(3)}/M^{(2)})\to H_3(M^{(3)}/M^{(2)})\,$ and 
$\,r_2:\,\pi_3(S(M^{(2)}/M^{(1)}))\to H_3(S(M^{(2)}/M^{(1)}))\,$ 
are the Hurewicz isomorphisms. Note, that the vertical 
arrows in (\ref{diag2}) are isomorphisms. Thus, 
$$
[M,\tilde G]\cong \hbox{coKer}(\partial^*\circ q^*)
= H^3(M; \pi_3(\tilde G))\cong H^3(M; H_3(\tilde G))\,.
$$
Since the fundamental group of a Lie group is abelian, 
$\,\pi_1(G_0)\cong H_1(G_0; \mathbb Z)$. It follows that 
$\,\hbox{Hom}(\pi_1(M),\pi_1(G_0))\cong 
\hbox{Hom}(H_1(M),H_1(G_0))$. 
The universal coefficient theorem, \cite{Br}, gives 
$$
0\to \hbox{Ext}^1_{\mathbb Z}(H_0(M),H_1(G_0))\to
H^1(M;H_1(G_0))\to \hbox{Hom}(H_1(M),H_1(G_0))\to 0\,.
$$
Since $\,\hbox{Ext}^*_{\mathbb Z}(\mathbb Z,\hbox{---})=0$, we have 
$\,\hbox{Hom}(H_1(M),H_1(G_0))\cong H^1(M; H_1(G_0))$. 
\medskip

Recall that the Skyrme functional is invariant under 
right multiplication by elements of $G$. We are therefore 
interested in the classification of maps from $M$ to $G$ 
up to homotopy and right translation in $G$. 
The following is a corollary of Proposition \ref{prop1}.
\begin{corollary}\label{corollary1}
$$
[M,\,G]/G\,\cong\,
H^3(M; H_3(\tilde G))\times H^1(M; H_1(G_0))\,.
$$
\end{corollary} 
\medskip

%%%%%%%%%%%%%%%%%

\subsection{Analytical description}
\label{sec:2.2}

We will now develop analytical expressions for the homotopy invariants 
starting with $\,H^3(M; H_3(\tilde G))$. 
Since the second homology of any 3-manifold is torsion free, 
the universal coefficient theorem gives 
$\,H^3(M;\,H_3(\tilde G))\,\cong\,\hbox{Hom}_{\mathbb Z}
(H_3(M),\,H_3(\tilde G))\,$. 
The universal covering group is the direct product of 
$\,\mathbb R^n\,$ together with a finite collection of compact, 
simply - connected, 
simple Lie groups, 
\begin{equation}\label{gtilde}
\tilde G\,\cong\,\mathbb R^n\times \tilde G^1\times\dots\times \tilde G^N\,.
\end{equation}
It is known, \cite{Bott}, that $\,\pi_3\,$ of any compact simple Lie group 
is $\,\mathbb Z$. Therefore, $\,H_3(\tilde G)\,$ is free, 
and, by the universal coefficient theorem, 
$\,H_3(\tilde G)\,\cong \,\hbox{Hom}_{\mathbb Z}
(H^3(\tilde G;\mathbb Z),\mathbb Z)$. 

We are interested in maps from $\,M\,$ to $\,G$, and not every 
map can be lifted to a map from $\,M\,$ to $\,\tilde G$. 
We would like to identify $\,H^3(\tilde G;\mathbb Z)\,$ 
with a subgroup of $\,H^3( G;\mathbb R)$ in order to express 
the homotopy $\,H^3$-invariant as the integral over $\,M\,$ of the pull-back 
of a 3-form. The needed identification is the topic of the next lemma.

Let $\,Z\,$ be the center of $\,G_0\,$ and $\,Z_0\,$ be the identity 
component of $\,Z$. Let $\,G^k\,$ be the connected Lie subgroup 
of $\,G_0\,$ corresponding to the Lie algebra of $\,\tilde G^k$. 
Let $\,d^k\,$ be the degree of the cover $\,\tilde G^k\to G^k$.  

%%%%%%%%%%%%%%%%%%%%%%%%%

\begin{lemma}\label{lemma1} 
The covering map, $\,p:\,\tilde G\to G$, induces 
a surjective homomorphism
$$
p^*:\,H^3(G_0;\mathbb R)\,\to\,H^3(\tilde G;\mathbb R)\,,
$$
which restricts to an isomorphism 
$$
\bigoplus_{k=1}^N H^3(G^k; \frac1{d^k} \mathbb Z) \to H^3(\tilde G;\mathbb Z)\,.
$$ 
\end{lemma}
\noindent{\bf Proof.} Consider the diagram:
\begin{equation}\label{diag3}
\begin{diagram}
p^{-1}(Z_0) & \rTo & \tilde G & \rTo^{j_Z} & \tilde G/p^{-1}(Z_0)\\
\dTo_p & & \dTo_p & & \dTo_q \\
Z_0& \rTo_{i_Z} &  G_0 & \rTo &  G_0/Z_0\\
\end{diagram}
\end{equation}
The homomorphism $\,i_{Z *}:\,\pi_1(Z_0)\to\pi_1(G_0)\,$ is injective, 
\cite{Brocker}, so $\,\pi_1(p^{-1}(Z_0))=1\,$ and 
$\,p^{-1}(Z_0)\cong \mathbb R^n$. 
Therefore, the homotopy exact sequence applied to the top row of (\ref{diag3}) 
implies that $\,\pi_m(\tilde G)\cong \pi_m(\tilde G/p^{-1}(Z_0))$. Thus, 
$\,H_3(\widetilde{(G/Z_0)})\cong 
\pi_3(\tilde G/p^{-1}(Z_0))\cong \pi_3(\tilde G)\cong H_3(\tilde G)$. 
Now, $\,G_0/Z_0\,$ being semisimple, has finite fundamental group. 
Define a transfer map $\,\tau:\,H_3(G_0/Z_0)\to H_3(\widetilde{(G_0/Z_0)})$ 
by sending every simplex to the sum of its lifts. The compositions 
$\,\tau\circ q_*\,$ and $\,q_*\circ \tau\,$ are multiplications by 
$\,|\pi_1(G_0/Z_0)|$, the order of the group. Thus, $\,p^*\,$ is surjective. 
The transfer map from $\,H^3(\tilde G^k; \mathbb R)\,$ to 
$\,H^3( G^k; \mathbb R)\,$  induces an isomorphism 
$\,H^3(\tilde G^k; \mathbb Z)\cong H^3( G^k; \frac{1}{d^k}\mathbb Z)$.  
The result now follows from the K\"unneth formula.
\medskip

For each $\,\tilde G^k\,$ in the decomposition (\ref{gtilde}), we have
a bi-invariant 3-form  on  $\,\tilde G^k\,$ defined by 

%%%%%%%%%%%%%%%%%%%%

$$
\tilde\Theta^k(X_g, Y_g, Z_g)\,=\,-\,\frac{K_{\tilde G^k}}{32\pi^2}
\,
\hbox{Tr}\,\left(\hbox{ad}\left[L_{g^{-1}*}X_g,\,L_{g^{-1}*}Y_g\right]
\hbox{ad}\,\left(L_{g^{-1}*}Z_g\right)\right)\,.
$$
The form $\,\tilde\Theta^k\,$ is closed because the Killing form is biinvariant.
Integrating it over any $\,SU(2)\,$ in $\,\tilde G^k$, we see that it 
is non-trivial. The constants $\,K_{\tilde G^k}\,$ are chosen so that 
the integral of this 3-form is 1 on any primitive $\,S^3\,$ in $\,\tilde G^k$. 
The value of this constant for every simple, simply connected, compact 
Lie group is listed in Table \ref{tab:1}. 

%%%%%%%%%%%%%%%%%

% For tables use
\begin{table}
%\caption{Please write your table caption here}
% For LaTeX tables use
\centering
\begin{tabular}{|l|l l l l l l l l l|}
\hline%\noalign{\smallskip}
group, $G$ & $A_n$ & $B_n$ & 
$C_n$ & $D_n$ & $E_6$ &
$E_7$ & $E_8$ & $F_4$ & $G_2$ \\ 
 & $SU(n+1)$ & $\hbox{Spin}(2n+1)$ & $\hbox{Sp}(n)$ & $\hbox{Spin}(2n)$ 
& & & & & \\
\hline% 
$K_G$   & 
 $\frac{2}{n+1}$ &
 $\frac{1}{2n-1}$ &
 $\frac{2}{n+1}$ & 
 $\frac{1}{2n-2}$ &
 $\frac{1}{24}$   &
 $\frac{1}{36}$   &
 $\frac{1}{240}$  &
 $\frac{1}{9}$ &
 $\frac{1}{2}$ \\ 
%\noalign{\smallskip}
\hline
\end{tabular}
% Or use
%\vspace*{5cm}  % with the correct table height 
\caption{Normalizing constants}\label{tab:1}  % Give a unique label
\end{table}

%%%%%%%%%%%%%%%%%%%%%%%%%%%%%%%

We now summarize how these constants are determined. 
The first observation is that $\,\pi_3\,$ of every 
simple, simply connected, compact 
Lie group is generated by a homomorphic image of 
$\,\hbox{Sp}(1)=SU(2)=\hbox{Spin}(3)$. To see this, one uses 
the homotopy exact sequence of the fibrations 
$$
U(n)\to U(n+1)\to S^{2n+1}\quad\hbox{and}\quad 
SU(n)\to U(n)\to S^1
$$
for $A_n$, 
$$
SO(m)\to SO(m+1)\to S^{m+1}
$$
for $B_n$ and $D_n$, and 
$$
\hbox{Sp}(n)\to \hbox{Sp}(n+1)\to S^{4n+3}
$$
for $C_n$. 

The exceptional groups are treated separately beginning with the 
inclusions (see \cite{Adams})
$$
\begin{array}{c}
\hbox{Spin}(3)\to \hbox{Spin}(9)\to F_4\to \\ 
\to \hbox{Spin}(10)\times_{\mathbb Z_4} S^1\to E_6 
\to \hbox{Spin}(12)\times_{\mathbb Z_2} \hbox{Sp}(1)
\to  E_7\to SO(16)\to E_8\,.
\end{array}
$$ 
Let $\,\gamma\,$ denote the generator of $\,\pi_3(\hbox{Spin}(3))$.
We have already seen that $\,\pi_3(\hbox{Spin}(9))\,$ and 
$\,\pi_3(\hbox{Spin}(10)\times_{\mathbb Z_4} S^1)\,$ are generated 
by $\,\gamma$. It follows that $\,\pi_3(F_4)\,$ is generated by 
$\,\gamma\,$ as well. Notice that 
$\,\hbox{Spin}(3)\,$ is naturally included in the 
$\,\hbox{Spin}(12)\,$ factor of 
$\,\hbox{Spin}(12)\times_{\mathbb Z_2} \hbox{Sp}(1)$. 
It follows that $\,\gamma\,$ is primitive in 
$\,\pi_3(\hbox{Spin}(12)\times_{\mathbb Z_2} \hbox{Sp}(1))$, thus 
$\,\pi_3(E_6)\,$ is generated by 
$\,\gamma$. Once again, $\,\gamma\,$ generates $\,\pi_3(SO(16))$, 
and, hence,  generates $\,\pi_3(E_7)$. Finally, 
the homotopy exact sequence of 
the fibration $\,SO(16)\to E_8\to E_8/SO(16)\,$ 
shows that $\,\gamma\,$ generates $\,\pi_3(E_8)\,$ since 
the perversely named octooctonionic projective plane 
$\,E_8/SO(16)\,$ is 4-connected, \cite{Mimura}, p.361. 
For the group $\,G_2\,$ we use the homotopy exact sequence of the fibration 
$\,SU(3)\to G_2\to S^6$. 

A straightforward evaluation of the integral 
$\,\int_{SU(2)}\tilde\Theta\,$ shows that $\,K_{SU(2)}=1$. 
For the other groups, one must integrate $\,\tilde\Theta\,$ over $\,\gamma$. 
Let $\,h:\,SU(2)\to\tilde G^k\,$ be a homomorphism generating 
$\,\pi_3(\tilde G^k)$. 
The constant is then the ratio of the value of the 
$\,SU(2)$-Killing form of a non-zero vector in $\,\frak{su}(2)$ 
to the value of the $\,\tilde G^k$-Killing form of its image under $\,h_*$. 
For the classical groups these ratios may be inferred from 
\cite{Simon} pp. 197, 199, 201, and 203, 
for example. For $\,E_6$, $\,E_7$, and $\,E_8$, these ratios may be found 
in \cite{Adams} p. 87, 77, and 43. We could not find references  
listing the constants for 
$\,F_4\,$ and $\,G_2$, so we compute them here. 

A nice description of 
$\,G_2\,$ may be found in \cite{Gross}.
By definition, $\,G_2\,$ is the group of endomorphisms of 
the purely imaginary octonions preserving the forms 
$\,\hbox{Re}(x^*y)\,$ and $\,\hbox{Re}([x,y]^*z)$. Its Lie 
algebra may be realized as the space of $7$ by $7$ matrices 
of the form
$$
\left[
\begin{array}{cccclccc}
0 & -\lambda_2 & -\lambda_3 &-\lambda_4 & &-\lambda_5 &-\lambda_6 & 
-\lambda_7 \\ 
\lambda_2 & 0 &-\mu_3 &-\mu_4 & &-\mu_5 &-\mu_6 & 
-\mu_7 \\ 
\lambda_3 & \mu_3 & 0 & \mu_5-\lambda_6 & & -\lambda_7-\mu_4 & 
\lambda_4-\mu_7 & \lambda_5+\mu_6 \\ 
\lambda_4 & \mu_4 & \lambda_6-\mu_5 & 0 & &  -\nu_5 & -\nu_6 & -\nu_7 \\
\lambda_5 & \mu_5 & \lambda_7+\mu_4 &  \nu_5 & & 0 & 
-\lambda_2-\nu_7 & \nu_6-\lambda_3 \\ 
\lambda_6 & \mu_6 & -\lambda_4+\mu_7 & \nu_6 & & \lambda_2+\nu_7 & 0 & 
-\mu_3 -\nu_5 \\ 
\lambda_7 & \mu_7 & -\lambda_5-\mu_6 & \nu_7 & & \lambda_3-\nu_6 & 
\mu_3+\nu_5 & 0 \\
\end{array}
\right]
$$
and the basis of the Lie algebra is obtained by setting 
each of the 14 parameters in turn to $1$ and the others to $0$. 
The Lie algebra of the $SU(3)$ mentioned above is obtained by setting 
all of the $\,\lambda$'s to $0$. The homomorphism $h$ 
induces a homomorphism of Lie algebras. The image of 
$$
v=\left[\begin{array}{cc} i & 0\\ 0 & -i\end{array}\right]\in \frak{su}(2)
$$
in the Lie algebra of $\,G_2\,$ is obtained by substituting $\,\nu_5=1\,$ 
and setting the rest of the parameters to $0$. The $14$ by $14$ matrix 
for $\,\hbox{ad}(h_*(v))\,$ in the given basis is obtained 
by direct computation, and the trace of its square is $\,(-16)$, thus, 
giving $\,K_{G_2}\,=\,-8/-16\,=\,1/2$. 

A matrix representation for $\,F_4\,$ would require $52$ by $52$ matrices.
Instead, we use generators and relations. 
Recall, that the $m$-dimensional Clifford algebra $\,CL_m\,$ has 
$m$ generators, $\,e_1$, \dots, $\,e_m$, with relations 
$\,e_i^2=-1$, and  $\,e_ie_j=-e_je_i$ for $i\neq j$. The Lie algebra 
$\,\frak{spin}(m)\,$ has  $\,e_ie_j\,$, $i<j$, as a basis, together 
with the usual Lie bracket, 
$\,[e_ie_j,\,e_ke_\ell] = e_ie_je_ke_\ell - e_ke_\ell e_ie_j$. 
Note, that the only nonzero brackets are of the form 
$\,[e_ie_j,\,e_ie_\ell]=2e_je_\ell
=-[e_ie_j,\,e_\ell e_i]\,$ when $\,j\neq \ell$. One may use this to 
recover the constants $K_{\hbox{Spin}(m)}$. 
We identify the $9$-dimensional spinors with the positive 
$10$-dimensional spinors, i.e.,
$$
\Delta^+_{10}=\Delta_9=
\{a\in CL^{\hbox{even}}_{10}\otimes \mathbb C\,|\,
e_{2j-1}e_{2j}\otimes i\cdot a\,=\,-a,\,j=1,2,3,4,5\}.
$$
Here $\,CL^{\hbox{even}}_{10}\,$ is the subspace of 
$\,CL_{10}\,$ generated by products of even numbers of $e_j$'s. 
Define $\epsilon_j=1-e_{2j-1}e_{2j}\otimes i\,$ and 
$\omega_j=e_{2j-1}+e_{2j}\otimes i$. Then 
$$
\begin{array}{llll}
\hskip-.1in\{\; \epsilon_1 \epsilon_2\epsilon_3\epsilon_4\epsilon_5,&
\omega_1 \omega_2\epsilon_3\epsilon_4\epsilon_5, &
\omega_1 \epsilon_2\omega_3\epsilon_4\epsilon_5,&
\omega_1 \epsilon_2\epsilon_3\omega_4\epsilon_5,\,\\
 \omega_1 \epsilon_2\epsilon_3\epsilon_4\omega_5,&
\epsilon_1 \omega_2\omega_3\epsilon_4\epsilon_5,&
\epsilon_1 \omega_2\epsilon_3\omega_4\epsilon_5,&
\epsilon_1 \omega_2\epsilon_3\epsilon_4\omega_5,\,\\
 \epsilon_1 \epsilon_2\omega_3\omega_4\epsilon_5,&
\epsilon_1 \epsilon_2\omega_3\epsilon_4\omega_5,&
\epsilon_1 \epsilon_2\epsilon_3\omega_4\omega_5,&
\omega_1 \omega_2\omega_3\omega_4\epsilon_5,\\
 \omega_1 \omega_2\omega_3\epsilon_4\omega_5,&
\omega_1 \omega_2\epsilon_3\omega_4\omega_5,&
\omega_1 \epsilon_2\omega_3\omega_4\omega_5,&
\epsilon_1 \omega_2\omega_3\omega_4\omega_5\;\}
\end{array}
$$
is a basis for $\,\Delta_9$. As a vector space, the Lie algebra 
$\,\frak{f}_4\,$ is $\,\frak{spin}(9)\oplus \Delta_9$. The Lie bracket 
of $\,\frak{f}_4\,$ restricted to $\,\frak{spin}(9)\,$ is the usual 
$\,\frak{spin}(9)$-bracket. The Lie bracket between the $\,\frak{spin}(9)\,$ 
and $\,\Delta_9\,$ factors is given by $\,[a,\,v]\,=\,a\, v$, the usual 
representation induced from Clifford multiplication. For the remaining 
brackets see \cite{Adams}, Lemma 6.2. The Lie algebra of the primitive 
$\,SU(2)\,$ in $\,\frak{f}_4\,$ is generated by 
$\,e_1e_2$, $\,e_1e_3$, and $\,e_2e_3$. A direct computation generates 
the $52$ by $52$ matrix for the $\,\frak{f}_4\,$ adjoint representation 
of $\,e_1e_2$. The trace of the square of this matrix is $-72$, giving 
$\,K_{F_4}\,=\,-8/-72\,=\,1/9$.  
\bigskip

Recall that $\, G^k\,$ is the connected subgroup of $\,G_0\,$ 
with universal covering group $\,\tilde G^k$.  Let $\,\Theta^k\,$ 
be the $3$-form on $\,G\,$ given by 
$$
\Theta^k(X_g, Y_g, Z_g)\,=\,-\,\frac{1}{32\pi^2}\,K_{\tilde G^k}\,
\hbox{Tr}\,\left(\hbox{ad}\left[\widehat{L_{g^{-1}*}X_g},
\,\widehat{L_{g^{-1}*}Y_g}\right]
\hbox{ad}\,\left(\widehat{L_{g^{-1}*}Z_g}\right)\right)\,, 
$$
where now $\,X_g$, $Y_g$, and $Z_g$ are tangent vectors on $\,G$, 
and $\,\widehat{L_{g^{-1}*}W_g}\,$ is the orthogonal projection of 
$\,L_{g^{-1}*}W_g\,$ onto the Lie algebra of $\,G^k$. Note, that  
since $\,\tilde G^k\,$ is a $\,d^k$-fold cover of $\,G^k$, 
the form $\,\Theta^k\,$ is not in general an integral class.  

Summarizing, we have the following lemma.

%%%%%%%%%%%%%%%%%%%

\begin{lemma}\label{lemma2}
The cohomology group, $\,H^3(\tilde G; \mathbb Z)$,  
is the free abelian group generated by 
$\,( \iota\circ p)^*\Theta^k$, where $\,p:\,\tilde G^k\to G_0\,$ is the 
covering projection, and $\,\iota:\,G_0\to G\,$ is the inclusion.
\end{lemma}
Before we can get numerical expressions for the three dimensional part 
of the homotopy invariants, we need to  
consider the second invariant 
that corresponds to $\,H^1(M; H_1( G_0))$, see Corollary \ref{corollary1}. 

Using the identification $\,H_1(G_0)\cong\pi_1(G_0)
\cong\{p^{-1}(1)\}\subset \tilde G$, a map $\,u:\,M\to G\,$ 
induces a $\,\tilde G$-connection $\,u^{-1} du$. 
Given a loop $\,\gamma:\,([0,1],\{0,1\})\to (M, x_0)$, solve 
the system of ordinary differential equations, 
$\,\frac{d\hfil}{dt} g^\gamma_t\,=\,u^{-1} du\,g^\gamma_t$, 
$\,g^\gamma_0=1$ to obtain a path $\,g^\gamma:\,[0,1]\to \tilde G$. 
The element of $\,H^1(M; H_1(G_0))\,$ corresponding to $u$ is 
$\,\alpha_u$, where $\,\alpha_u([\gamma])= g^\gamma_1$. 
This is just the holonomy of the connection. We take up the full description 
of holonomy in Section 3. 

Note, that $\,H^1(M; H_1( G_0))\,$ is a finitely generated abelian 
group. Let $\,\alpha_1,\dots,$ $\dots,\alpha_b\,$ be its generators, and  
$\,r_\ell\,\alpha_\ell=0\,$ be the relations, where 
$\,r_1\,|\,r_2\,|\dots\,|\,r_b$. There is no canonical choice 
for the classes $\,\alpha_\ell$, but it is not difficult to 
construct a set given any closed connected $3$-manifold. 
By the proof of Proposition \ref{prop1}, there are maps 
$\,v_\ell:\,M\to G_0\,$ with 
$\,(v_\ell)_*=\alpha_\ell\,:\,H_1(M)\to H_1(G_0)$. 
It is usually possible to construct such maps explicitly. 

Given a map $\,u:\,M\to G$, one obtains the map 
$\,u_0\,=\,u(x_0)^{-1}u\,$ form $\,M\,$ into $\,G_0$. 
This map induces a map $\,u_{0*}:\,H_1(M)\to H_1(G_0)$. 
We write $\,u_{0*}=\sum_{\ell=1}^b a_\ell\,\alpha_\ell$, 
where $\,a_\ell=a_\ell(u)\in\mathbb Z_{r_\ell}$. 

We are in position now to state the analytical description of 
$\,[M,\,G]/G$. 

%%%%%%%%%%%%%%%%

\begin{proposition}\label{prop2} 
Given any element $\,[u]\,$ of $\,[M,\,G]/G$, define the numbers 
$\,a_\ell(u)\,$ and 
$$
c^k(u)\,=\,\int_M (\,u(\prod_1^b v_\ell^{a_\ell})^{-1}\,)^*
\Theta^k\,.
$$ 
We have $\,a_\ell(u)\in\mathbb Z_{r_\ell}$ and $\,c^k(u)\in \mathbb Z$.
Any $\,(b+N)$-tuple 
$$
(a_1,\dots,a_b;\,c^1,\dots,c^N)\,\in\,\mathbb Z_{r_1}\times\dots\times
\mathbb Z_{r_b}\times \mathbb Z^N\,,
$$
is obtained from some map $\,u:\,M\to G$. Two maps are equivalent if 
and only if they produce the same $\,(b+N)$-tuple. 
\end{proposition}

%%%%%%%%%%%%%%%%%

\begin{remark}\label{valpha} The three dimensional part of this invariant is a 
torsor, i.e., an affine space modeled on integers with no canonical 
choice of $0$. Since
$$
\int_M (u w)^*\Theta^k\,=\,\int_M u^*\Theta^k\,+\,\int_M w^*\Theta^k\,,
$$
we could encode the information contained in $\,c^k(u)\,$ as 
$\,\int_M u^*\Theta^k$. However, these numbers are no longer integers, 
but the differences $\,\int_M u^*\Theta^k\,-\,\int_M w^*\Theta^k$ 
always are.  
\end{remark} 
\begin{remark}\label{r2}
 The integrals $\,\int_M u^*\Theta^k\,$ are well 
defined for maps $\,u:\,M\to G\,$ with finite Skyrme energy, $\,E(u)$, 
since $\,u^{-1} du\in L^2\,$ and $\,[u^{-1} du,\,u^{-1} du]\in L^2$. 
In Section 4 we will see that all of the numbers $\,a_\ell\,$ and 
$c^k(u)\,$ are well defined for maps with finite Skyrme energy.
\end{remark}

%%%%%%%%%%%%%%%%%%%%%%%  SECTION 3 %%%%%%%%%%%%%%%%%%%%%%%%%%%%%%%%

\section{Connections and holonomy}\label{sec:3}

In the previous section we described the one dimensional component 
of the homotopy invariant using the solutions of a system of ordinary 
differential equations. While this is legitimate for smooth maps, 
it does not directly apply to all maps with finite Skyrme energy. 
To circumvent this problem, we will give a new definition of holonomy 
in the spirit of \v Cech cohomology that is well defined for 
distributionally flat $\,L^2$-connections and reduces to the usual 
definition for smooth connections.

%%%%%%%%%%%%%%%%%%%%%%%%%%%%

\subsection{Function spaces}\label{sec:3.1} 
Every compact Lie group has a faithful unitary representation, \cite{Brocker}. 
Pick such a representation and identify $\,G\,$ with its image, 
a subgroup of $\,U(N)\subset \mathbb C^{N^2}$. Use the corresponding 
representation to identify the Lie algebra of $G$, $\,\frak g$, 
with a subalgebra of $\,\frak{u}(N)\subset \mathbb C^{N^2}$. 
These identifications allow us to consider $\,G\,$ and $\,\frak g\,$ valued 
maps as maps into $\,\mathbb C^{N^2}$.

Given a Riemannian manifold $\,\Omega$, denote by 
$\,W^{s,p}(\Omega,\mathbb C^m)\,$ the Sobolev space of $\,\mathbb C^m$-valued 
functions on $\,\Omega$. We use these spaces only when $\,s\,$ 
is nonnegative integer, in which case the norm can be chosen as 
$$
\|u\|_{W^{s,p}(\Omega, \mathbb C^m)}^p\,=\,
\int_\Omega\sum_{0\le k\le s}|\nabla^s u|^p\,d\,\hbox{vol}_\Omega\,.
$$
We define $\,W^{s,p}(\Omega,G)\,$ to be the space of maps 
$\,u\in W^{s,p}(\Omega,\mathbb C^{N^2})\,$ with $\,u(x)\in G\,$ 
for almost all $\,x\in\Omega$. We define $\,W^{s,p}(\Omega,\frak g)\,$ 
similarly. Notice, that 
$\,W^{s,p}(\Omega,G)$ $\subset L^\infty(\Omega,\mathbb C^{N^2})$, since $\,G\,$ 
is compact. 
\begin{remark}\label{remark3} The space $\,W^{1,p}(\Omega, G)\,$ forms a group 
under pointwise multiplication.
\end{remark}

We will sometimes shorten the notation to $\,W^{s,p}\,$  
and will use the same notation for the corresponding spaces of 
differential forms on $\,\Omega$. Different faithful representations 
of $\,G\,$ lead to equivalent norms. For compact manifolds, 
different Riemannian metrics also lead to equivalent norms 
provided the boundary is sufficiently regular.

%%%%%%%%%%%%%%%%%%%%%%%%%%%

\subsection{Nonlinear Poincar\'e Lemma}\label{sec:3.2} 
In this section we prove an important lemma that we use to 
define the holonomy for Sobolev connections. 

Let $\,I^m\,$ denote a unit cube in $\,\mathbb R^m$. 
\begin{lemma}\label{poinc}
Given any $\,L^2\,$ $\,\frak g$-valued $1$-form $\,A\,$ on $\,I^m\,$ 
such that 
\begin{equation}\label{flat1}
dA\,+\,\frac{1}{2}\,[A,\,A]\,=\,0
\end{equation} 
in the sense of distributions, there exists 
$\,u\in W^{1,2}(I^m,\,G)\,$ such that $\,u^{-1}\in W^{1,2}(I^m,\,G)\,$ 
and $\,A\,=\,u^{-1}\,du$. 
Furthermore, for any two such maps, $\,u\,$ and $\,v$, there exists 
$\,g\in G\,$ so that  $\,u(x)\,=\,g\cdot v(x)$, for almost every $\,x\in I^m$.
\end{lemma}
\noindent{\bf Proof.\/} Choose coordinates parallel to the edges of 
the cube. 
In coordinates, the 
$\,\frak g$-valued $1$-form $\,A\,$ can be written as $\,A_k(x)\,dx^k$, 
where each $\,A_i(x)\,$ is a matrix-valued function. We then have 
$\,[A,\,A]\,=\,(A_iA_j\,-\,A_jA_i)\,dx^i\wedge dx^j$. Here and in what follows 
we use the summation convention.  

%%%%%%%%%%%%%%%

The following observation is important for 
our construction. Let $\,f\,$ be a scalar function in $\,L^p(I^m)$, for some 
$\,1\le p<\infty$. By Fubini's theorem, for almost all values of 
$\,x^1\in I^1\,$ the function of $\,m-1\,$ variables $\,f(x^1,\cdot)\,$ 
is in $\,L^p(I^{m-1})$. 
We will extend $\,f\,$ outside of $\,I^m\,$ by $\,0$.
Denote by $\,T_\epsilon f\,$ the mollification of $\,f\,$ 
defined as follows:
$$
(T_\epsilon f)(x^1,\dots,x^m)\,=\,\int 
\zeta_\epsilon(x^1-y^1)\cdot\dots\cdot \zeta_\epsilon(x^m-y^m)f(y^1,\dots, y^m)\,
d^m y\;,
$$ 
where $\,\zeta_\epsilon(t)=\epsilon^{-1}\zeta(\epsilon^{-1} t)\,$ with 
$\,\zeta\,$ a smooth, even, compactly supported  
bump-function with integral $1$. Choose a sequence $\,\epsilon_k\to 0\,$ so that 
$$
\sum_{k=1}^\infty \int_{I^m}|T_{\epsilon_k} f\,-\,f|^p\,d^m x\,<\,\infty\,.
$$
Then, 
$$
\int_{I^1}\left(\sum_{k=1}^\infty \int_{I^{m-1}}
|T_{\epsilon_k} f(x^1,\cdot)\,-\,f(x^1,\cdot)|^p\,dx^2\dots dx^m\right)\,dx^1\,
<\,\infty\,.
$$
This implies that there is a subset $\,\tilde I^1\,$ of full measure in 
$\,I^1\,$ for each point $\,x^1\,$ of which $\,T_{\epsilon_k}f(x^1,\cdot)\,$ 
converges to $\,f(x^1,\cdot)\,$ in $\,L^p(I^{m-1})$. If there is a finite 
number of functions $\,f$, the set $\,\tilde I^1\,$ can be chosen to 
accomodate all of them.

%%%%%%%%%%%%%%%%%%

We now return to the connection $\,A\in L^2(I^m, G)$. Using 
the above observation we
translate the coordinates in $\,I^m\,$ as follows. 
For almost every point $\,x^1_0\,$ of the interval $\,I^1$, 
the restrictions of all $\,A_i\,$ to the hyperplane $\,x^1=x^1_0\,$ 
lie in $\,L^2(I^{m-1})\,$ and  
$\,(A_i)_{\epsilon_k}(x^1_0,\cdot) = (T_{\epsilon_k}A_i)(x^1_0,\cdot)\,$ 
converge to $\,A_i(x^1_0,\cdot)\,$ in $\,L^2(I^{m-1})$. 
In addition, for almost every such $\,x^1_0$,
the restrictions of all Lie brakets $\,[A_j, A_\ell]\,$ 
to the hyperplane $\,x^1=x^1_0\,$ 
lie in $\,L^1(I^{m-1})\,$ and  
$\,([A_j, A_\ell])_{\epsilon_k}(x^1_0,\cdot) = 
(T_{\epsilon_k}[A_j, A_\ell])(x^1_0,\cdot)\,$ 
converge to $\,[A_j, A_\ell](x^1_0,\cdot)\,$ in $\,L^1(I^{m-1})$. 
Fix one such point $\,x^1_0\,$ inside $\,I^1$. Similarly 
(sparsing the sequence $\,\epsilon_k$, if necessary)
we may fix values $\,x^2_0$,\dots, $\,x^m_0\,$ so that 
the restrictions of all $\,A_i\,$ to the 
slices $\,\{x^1_0\}\times\dots\times\{x^k_0\}\times I^{m-k}\,$ 
are in $\,L^2(I^{m-k})\,$ and the mollifications 
$\,\int \zeta_\epsilon(x^k_0-y^k)A_i(x^1_0,\dots,x^{k-1}_0, y^k,\cdot)\,dy^k\,$ 
converge to $\,A_i(x^1_0,\dots,x^{k-1}_0, x^k_0,\cdot)\,$ 
in $\,L^2(I^{m-k})$. In addition, the restirctions 
of all $\,[A_j, A_\ell]\,$ to the 
slices $\,\{x^1_0\}\times\dots\times\{x^k_0\}\times I^{m-k}\,$ 
are in $\,L^1(I^{m-k})\,$ and the mollifications 
$\,\int \zeta_\epsilon(x^k_0-y^k)[A_j, A_\ell]
(x^1_0,\dots,x^{k-1}_0, y^k,\cdot)\,dy^k\,$ 
converge to the braket 
$\,[A_j, A_\ell](x^1_0,\dots,x^{k-1}_0, x^k_0,\cdot)\,$ 
in $\,L^1(I^{m-k})$. 
Almost every point in $\,I^m$ satisfies these ``slice"  
conditions. By translating the coordinates we set $\,x_0=0$. 

After these preliminary remarks we turn to the construction of $\,u$. 
We obtain $\,u(x)\,$ as $\,u_n(x)$, where 
$\,u_k:\,\{0\}\times I^k\to G\,$ is defined inductively, 
setting $\,u_{0}(0)=\mathbf 1\,$ and 
defining successive terms 
via the system of ordinary differential equations 
\begin{equation}\label{ode}
\begin{array}{c}
\frac{d\hfil}{dt} v(t,x^{n-k},\dots, x^n)
=v(t,x^{n-k},\dots, x^n)
A_{n-k-1}(0,t,x^{n-k},\dots, x^{n}) \\ 
   v(0,\,x^{n-k},\dots, x^{n})\,=\,u_{k}(x^{n-k},\dots, x^{n})\;,
\end{array}
\end{equation}
setting $\,u_{k+1}(x^{n-k-1},x^{n-k},\dots, x^n)
=v(x^{n-k-1},x^{n-k},\dots, x^n)$. 
We will prove that system (\ref{ode}) 
has a unique solution in $\,L^2\,$ 
with $\,\frac{dv}{dt}\,$ also in $\,L^2$. It is not hard to see that, 
for any $\,i\ge n-k$, the quantity 
$\,w_i\,=\,\partial_i u_{k+1}\, -\, u_{k+1}\,A_i\,$ formally satisfies the 
differential equation
$$
\partial_j w_i = \partial_i(\partial_j u_{k+1} - u_{k+1} A_j)+
w_i A_j  
+ \,u_{k+1}
(\partial_i A_j - \partial_j A_i + A_i A_j - A_j A_i)\;,
$$
on the slice $\,\{0\}\times I^{k+1}\,$ with $\,j=n-k-1$. 
Using the defining differential equation for $\,u_{k+1}\,$ 
and hypothesis (\ref{flat1}), one obtains 
$\,\partial_j w_i = w_i A_j$. The initial condition $\,w_i(x_j=0)=0\,$ 
follows from the induction step. So, formally, $\,w_i=0$, which implies 
that $\,u_{k+1}\in W^{1,2}$. We now make this argument precise.

The initial value problem 

%%%%%%%%%%%

\begin{equation}\label{ode1} 
\frac{d\hfil}{dt} v(t)\,=\,v(t)\,a(t)\;,\qquad v(0)\,=\,v_0\in G\,,
\end{equation}
with $\,a\in L^2(I^1,\frak g)$, has a unique solution 
$\,v\in W^{1,2}(I^1, G)$. To prove this, we mollify $a$ to get 
$\,a_\epsilon(t)\,=\,(\zeta_\epsilon * a)(t)$ and solve the smooth system 

%%%%%%%%%%%%%%%

\begin{equation}\label{ode2} 
\frac{d\hfil}{dt} v_\epsilon(t)\,
=\,v_\epsilon(t)\,a_\epsilon(t)\;,\qquad v_\epsilon(0)\,=\,v_0\,,
\end{equation}
obtaining smooth functions $\,v_\epsilon\,$ with values in $\,G$. 
Since $\,G\,$ is compact, the functions $\,v_\epsilon\,$ 
are uniformly bounded, and they are equicontinuous on $\,I^1$, 
since
$$
\begin{array}{c}
|v_\epsilon(t_2)-v_\epsilon(t_1)|\,
=\,|\int_{t_1}^{t_2}v_\epsilon(s)\,a_\epsilon(s)\,ds|
\le C_1\,\int_{t_1}^{t_2}|a_\epsilon(s)|\,ds\\ 
\le\,C_1\,|t_2-t_1|^{\frac12}
\left(\int_{t_1}^{t_2}|a_\epsilon(s)|^2\,ds\right)^{\frac12}\,
\le\,C_1\,\|a\|_{L^2( I^1;\frak g)}\,|t_2-t_1|^{\frac12} 
\end{array}
$$ 
Hence, there is a sequence $\,\epsilon_k\to 0\,$ for which 
$\,v_{\epsilon_k}\,$  converges uniformly on $\,I^1$  
to a continuous $\,G$-valued function $\,v$. Also, 
$\,v_\epsilon\,a_\epsilon  \to v\,a\,$ in $\,L^2$. This implies that 
$\,{d\hfil\over dt} v_{\epsilon_k}\,$ converges in $\,L^2$; the limit is 
the distributional derivative of $\,v$. This shows that 
$\,v\in W^{1,2}( I^1; G)$, and that $\,v\,$ is a solution of (\ref{ode1}). 
Note, that for all $\,t$ we have 

%%%%%%%%%%%%%%
 
\begin{equation}\label{odeint}
v(t)\,=\,v_0\,+\,\int_0^t v(s) a(s)\,ds\,,  
\end{equation}
and this equation is equivalent to (\ref{ode1}). If 
$\,w\in W^{1,2}(I^1; G)\,$ is another solution 
of (\ref{ode2}), then the difference, 
$\,p(t)=v(t)-w(t)$, satisfies
$\,
|p(t)|\,\le\,\int_0^t |p(s)|\,|a(s)|\,ds\, 
$,
and is therefore identically zero. 

This argument establishes the base case of the induction, namely, 
$\,u_1(0,\cdot)\in W^{1,2}(I^1, G)$. To complete the induction, 
we will prove that 
$\,u_{k+1}\,A_\ell\big|_{\{0\}\times I^{k+1}}$ 
$=\partial_\ell u_{k+1}\,$ 
for $\,\ell\ge n-k-1\,$ and 
$\,u_{k+1}\in W^{1,2}(\{0\}\times I^{k+1}, G)\,$ 
when 
$\,\partial_i u_{k}=u_{k}\,A_i\big|_{\{0\}\times I^{k}}\,$ 
for $\,i\ge n-k\,$ and
$\,u_{k}\in W^{1,2}(\{0\}\times I^{k}, G)$. 
We first need to show that the equation 
\begin{equation}\label{flat2}
dA\,\big|_{\{0\}\times I^{k+1}}\,
+\,\frac12\,[A\,\big|_{\{0\}\times I^{k+1}},\,A\,\big|_{\{0\}\times I^{k+1}}]
\,=\,0\,
\end{equation}
holds in the sense of distributions. We prove this inductively. 
By the hypothesis of the lemma, 
we have 

%%%%%%%%%%%%%%%%

\begin{equation}\label{flat3}
\int_{I^n} (A_j\,\partial_i\eta\,-\,A_i\,\partial_j\eta)\;d^n x\,=\,
\int_{I^n} [A_i,\,A_j]\,\eta\;d^n x
\end{equation}
for any smooth scalar function $\,\eta\,$ supported in 
the interior of the cube. When both $\,i\,$ and $\,j\,$ are larger than 
$\,1$, equation (\ref{flat3}) will also hold for scalar functions 
of the form  
$$
\eta_\epsilon(x^1,\dots, x^n)\,=\,
\int \zeta_\epsilon(-x^1)\zeta_\epsilon(y^2-x^2)\dots \zeta_\epsilon(y^n-x^n) 
\,\phi(y^{2},\dots,y^n)\,d^ny\,. 
$$
Substitute this $\,\eta_\epsilon\,$ in (\ref{flat3}), integrate by parts 
to move all derivatives 
onto $\,\phi$, and change the order of integration to obtain
$$
\begin{array}{c}
\int_{I^n} (A_j)_\epsilon(0,y^2,\dots,y^n)\,\partial_{y^i}\phi\,
-\,(A_i)_\epsilon(0,y^2,\dots,y^n)\,\partial_{y^j}\phi\;d^n y \\ 
 =\,
\int_{I^n} ([A_i,\,A_j])_\epsilon(0,y^2,\dots,y^n)\,\phi\;d^n y\,.
\end{array}
$$
Pass to the limit 
as $\,\epsilon\,$ goes to $\,0\,$ along the sequence chosen in the beginning 
of the proof. The result will be  
$$
\int_{I^{n-1}} (A_j(0,\cdot)\,\partial_i\phi\,
-\,A_i(0,\cdot)\,\partial_j\phi)\;d^{n-1} x\,=\,
\int_{I^{n-1}} [A_i(0,\cdot),\,A_j(0,\cdot)]\,\phi\;d^{n-1}x\,.
$$ 
Repeating this argument, in a finite number of steps we obtain 
(\ref{flat2}). 

We next consider a mollified version of system (\ref{ode}), 

%%%%%%%%%%%%%%%%%%%%

\begin{equation}\label{ode11}
\begin{array}{c}
\frac{d\hfil}{dt} v^\epsilon\,
=\,v^\epsilon
(A_{n-k-1}\big|_{\{0\}\times I^{k+1}})_\epsilon(t,x^{n-k},\dots, x^{n}) \,,
\\   
v^\epsilon(0)\,=\,(u_{k})_\epsilon (x^{n-k},\dots, x^{n})\;.
\end{array}
\end{equation}

Let $\,\Phi^\epsilon(t,s;\,x^{n-k},\dots, x^n)\,$ be  
the solution of 

%%%%%%%%%%%%%%%%%%%%%

%\begin{equation}\label{fund}
$$
\frac{d\hfil}{dt}\Phi^\epsilon(t,s;\cdot)\,=\,
\Phi^\epsilon(t,s;\cdot)\,
(A_{n-k-1}\big|_{\{0\}\times I^{k+1}})_\epsilon(t,\cdot)\,,
\qquad \Phi^\epsilon(s,s;\cdot)\,=\,\mathbf 1\,.
$$
%\end{equation}
In terms of $\,\Phi^\epsilon$, the solution of (\ref{ode11}) can be written 
as 
$$
v^\epsilon(x^{n-k-1},\dots, x^n)=
u_{k}(x^{n-k},\dots,x^n)\,\Phi^\epsilon(x^{n-k-1},0;\,x^{n-k},\dots, x^n)\,,
$$
and it is a smooth function of all of the variables. 
For any $\,i\ge n-k$, set 
$\,w_i^\epsilon\,=\,\partial_i v^\epsilon\, 
-\, v^\epsilon\,(A_i\big|_{\{0\}\times I^{k+1}})_\epsilon\,$ 
on $\,\{0\}\times I^{k+1}$. 
Differentiate equation (\ref{ode11})  
to obtain (with $\,j=n-k-1$)

%%%%%%%%%%%%%%%%%%%

$$
\begin{array}{c}
\partial_j w_i^\epsilon =  
\partial_i(\partial_j v^\epsilon 
- v^\epsilon (A_j\big|_{\{0\}\times I^{k+1}})_\epsilon) 
+
w_i^\epsilon (A_j\big|_{\{0\}\times I^{k+1}})_\epsilon \hskip1.5in\\ 
+ \,v^\epsilon
\left(\partial_i (A_j\big|_{\{0\}\times I^{k+1}})_\epsilon 
- \partial_j (A_i\big|_{\{0\}\times I^{k+1}})_\epsilon\right. \\
\left.+ (A_i\big|_{\{0\}\times I^{k+1}})_\epsilon 
(A_j\big|_{\{0\}\times I^{k+1}})_\epsilon 
- (A_j\big|_{\{0\}\times I^{k+1}})_\epsilon 
(A_i\big|_{\{0\}\times I^{k+1}})_\epsilon\right)\;,
\end{array}
$$
on the slice $\,\{0\}\times I^{k+1}\,$. 
Notice that $\,\partial_j v^\epsilon 
- v^\epsilon (A_j\big|_{\{0\}\times I^{k+1}})_\epsilon\,=\,0\,$ 
by (\ref{ode11}) and 

%%%%%%%%%%%%%%%%

\begin{equation}\label{F}
\begin{array}{rl}
F_{ij}^\epsilon\,:= & \partial_i (A_j\big|_{\{0\}\times I^{k+1}})_\epsilon 
- \partial_j (A_i\big|_{\{0\}\times I^{k+1}})_\epsilon \\ 
 & + (A_i\big|_{\{0\}\times I^{k+1}})_\epsilon 
(A_j\big|_{\{0\}\times I^{k+1}})_\epsilon 
- (A_j\big|_{\{0\}\times I^{k+1}})_\epsilon 
(A_i\big|_{\{0\}\times I^{k+1}})_\epsilon \\ 
= & (A_i\big|_{\{0\}\times I^{k+1}})_\epsilon 
(A_j\big|_{\{0\}\times I^{k+1}})_\epsilon 
- (A_j\big|_{\{0\}\times I^{k+1}})_\epsilon 
(A_i\big|_{\{0\}\times I^{k+1}})_\epsilon \\ 
& -
\left((A_i A_j - A_j A_i)\big|_{\{0\}\times I^{k+1}}\right)_\epsilon \,,
\end{array}
\end{equation}
by equation (\ref{flat2}). Also,
$$ 
\begin{array}{rl}
w_i^\epsilon(0,x^{n-k},\dots, x^n)\,& =\,\partial_i(u_k)_\epsilon - 
(u_k)_\epsilon \,(A_i\big|_{\{0\}\times I^{k+1}})_\epsilon \\ 
 & =\,
(\partial_i\,u_k-u_k\,A_i\big|_{\{0\}\times I^{k}})_\epsilon\,\\ 
& +\,
(u_k\,A_i\big|_{\{0\}\times I^{k}})_\epsilon\,
-\,(u_k)_\epsilon \,(A_i\big|_{\{0\}\times I^{k}})_\epsilon\\ 
& +\,(u_k)_\epsilon\,((A_i\big|_{\{0\}\times I^{k}})_\epsilon 
-(A_i\big|_{\{0\}\times I^{k+1}})_\epsilon)\\
 & =\,(u_k\,A_i\big|_{\{0\}\times I^{k}})_\epsilon\,
-\,(u_k)_\epsilon \,(A_i\big|_{\{0\}\times I^{k}})_\epsilon\\ 
& +\,(u_k)_\epsilon\,((A_i\big|_{\{0\}\times I^{k}})_\epsilon 
-(A_i\big|_{\{0\}\times I^{k+1}})_\epsilon)=: w^\epsilon_{i0}
,
\end{array}
$$
by the induction hypothesis. Thus, 
$\,w_i^\epsilon(x^{n-k-1},x^{n-k},\dots, x^n)\,$ 
is the solution of the following initial value problem:
$$
\partial_j\,w_i^\epsilon\,=\,w_i^\epsilon\,(A_j)_\epsilon\,
+\,v^\epsilon\,F_{ij}^\epsilon\,,\qquad  
w_i^\epsilon|_{x^j=0}\,=\,w^\epsilon_{i0}\;.
$$
Therefore, 
$$
w_i^\epsilon(x^j,\cdot)\,=\,
w^\epsilon_{i0}\,
\Phi^\epsilon (x^j,0;\cdot)\, 
+\,\int_0^{x^j}
(v_\epsilon\,F_{ij}^\epsilon)|_{x=(0,s,\cdot)}\,
\Phi^\epsilon(x^j,s;\cdot)\,ds\;.
$$ 
Notice that $\,w^\epsilon_{i0}\to 0\,$ in $\,L^1(\{0\}\times I^k)$,   
since $\,u_k\in W^{1,2}(\{0\}\times I^k, G)$, and 
$\,(A_i\big|_{\{0\}\times I^{k}})_\epsilon\to 
A_i\big|_{\{0\}\times I^{k}}\,$ in $\,L^2(\{0\}\times I^k,\frak g)$, 
and, due to the choice of the coordinates at the beginning of the argument, 
$\,(A_i\big|_{\{0\}\times I^{k+1}})_\epsilon\big|_{x^j=0}\to 
A_i\big|_{\{0\}\times I^{k}}\,$ 
in $\,L^2(\{0\}\times I^k, \frak g)$.  In addition, 
$\,\Phi^\epsilon\,$ is smooth 
and uniformly bounded, so 
$$
\| \,w^\epsilon_{i0}\,
\Phi^\epsilon (x^j,0;\cdot)\|_{L^1(\{0\}\times I^{k+1})}\,
\stackrel{\epsilon\to 0}{\rightarrow}\,0\,.
$$
Since $\,(A_\ell)_\epsilon\to A_\ell\,$ in $\,L^2(\{0\}\times I^{k+1})$, we have 
$\,F_{ij}^\epsilon\to 0\,$ in $\,L^1\,$ (see equation (\ref{F})). 
Both  $\,v^\epsilon\,$ and $\,\Phi^\epsilon\,$ are 
smooth and uniformly bounded, thus $\,w^\epsilon_i\,$ 
tends to $0$ in $\,L^1(\{0\}\times I^{k+1},\mathbb C^{N^2})$, i.e.,
\begin{equation}\label{limit1} 
\|\partial_i v^\epsilon\,
-\,v^\epsilon\,(A_i\big|_{\{0\}\times I^{k+1}})_\epsilon\,
\|_{L^1(\{0\}\times I^{k+1},\mathbb C^{N^2})}\,\to \,0\,.
\end{equation}  
Because the $\,L^1$-norm of 
$\,v^\epsilon\,(A_i\big|_{\{0\}\times I^{k+1}})_\epsilon\,$ is 
uniformly bounded, (\ref{limit1}) implies that 
the $\,L^1$-norm of $\,\partial_i v^\epsilon\,$ is bounded, while equation 
(\ref{ode11}) shows that $\,\partial_j v^\epsilon\,$ is bounded in $\,L^1\,$ 
as well. Thus, $\,v^\epsilon\,$ is bounded in $\,W^{1,1}(I^{k+1},G)$. 
Since in our case the embedding $\,W^{1,1}\to L^1\,$ is compact, 
there exists a sequence $\,\epsilon_m\to 0\,$ so that 
$\,v^{\epsilon_m}\,$ converges strongly in $\,L^1\,$ and almost everywhere 
in $\,I^{k+1}\,$ to some $\,v\in L^1(I^{k+1},G)$. In addition, 
we have $\,\partial_i v\,=\,v\,A_i(0,\cdot)\,$ in the sense of distributions 
for all $\,n-k-1\le i \le n$. Hence, $\,v\in W^{1,2}(I^{k+1},G)$. We still need 
to check the initial conditions in (\ref{ode}). Taking a further subsequence 
if necessary, we will have $\,v^{\epsilon_m}(t,\cdot)\to v(t,\cdot)\,$ in 
$\,L^1(I^k)\,$ for almost every $\,t\in I^1$. Indeed, let $\,\epsilon_m\,$ 
be such that $\,\|v^{\epsilon_m}-v\|_{L^1(I^{k+1})}\le 2^{-m}$. Then, 
$$
\int_{I^1}(\sum_{m=1}^\infty\int_{I^k}|v^{\epsilon_m}-v|\,d^kx)\,dt\,=\,
\sum_{m=1}^\infty\int_{I^{k+1}}|v^{\epsilon_m}-v|\,d^{k+1}x \le 1\,,
$$ 
which proves the convergence almost everywhere.  
To simplify the notation 
we will omit the subscript $\,m\,$ on the subsequence in the future. 
Multiply equation (\ref{ode11}) by
an arbitrary $\,\eta\in L^2(I^k)\,$ 
and an arbitrary smooth $\,\varphi\,$ and integrate to obtain 

%%%%%%%%%%

$$
\begin{array}{rl}
\varphi(t)(v^\epsilon(t),\eta) & = \varphi(0)(v^\epsilon(0),\eta)\\
&+\int_0^t ((v^\epsilon(s),\eta)\varphi^\prime(s)+
(v^\epsilon(s)(A_j)_\epsilon(0,s),\eta)\varphi(s))\,ds,
\end{array}
$$ 
where $\,(\xi,\eta)
=\int_{I^k}\xi\,\eta\,d^kx$. Each of the terms on the right hand side converge, 
as $\,\epsilon\to 0$, to the corresponding term with $\,v\,$. Hence, 
the left side must have a limit as well. For almost all $\,t\in I^1\,$ this 
limit must be $\,\varphi(t)(v(t),\eta)$. As a result, $\,v\,$ is a weakly 
continuous function from $\,I^1\,$ to $\,L^2(I^k)$. It satisfies the 
differential equation 
$$
\frac{d\hfil}{dt}\,(v(t),\eta)\,=\,(v(t)\,A_j(0,t),\eta)
$$
almost everywhere, and $\,v(t)\to v(0)=u_k\,$ weakly in $\,L^2(I^k)$. 
The fact, that $\,v\,$ is the unique solution to equation (\ref{ode}), 
follows from the argument given in the  base case. 
Thus, we have $\,du\,=\,u\,A$. Recall, that $\,u(x)\in G\subset U(N)\,$
for almost every $x$. Hence, $\,u^{-1}=u^*\,$ is in $\,W^{1,2}(I^n,G)$.
This concludes the 
proof of the existence of $\,u\in W^{1,2}(I^n,G)\,$ with 
$\,u^{-1}\,du = A$. 

Now assume that $\,\tilde u^{-1}\,d\tilde u = u^{-1}\,du\,$ for some 
other $\,\tilde u\in W^{1,2}(I^n,G)$. Since $\,\tilde u\,u^{-1}\,$ belongs 
to $\,W^{1,2}(I^n,G)$, we compute: $\,d(\tilde u\,u^{-1}) = 
d\tilde u\,u^{-1}\,-\,\tilde u u^{-1}\,du u^{-1}\,=\,
\tilde u\,(\tilde u^{-1}d\tilde u - u^{-1}d u)\,u^{-1} = 0$. Hence, 
$\,\tilde u(x)\,=\,g\,u(x)\,$ for some constant $\,g\in G$. 
Now the lemma is proved.

%%%%%%%%%%%%%%%%%

\subsection{The holonomy representation for $\,L^2_{loc}\,$ connections} 
Let $\,\Omega\,$ be an $\,n$ - dimensional Riemannian manifold. 
Let $\,P\,$ be a principal $\,G$-bundle over $\,\Omega$. 
Let $\,\{\varphi_\nu:\,{\cal U}_\nu\times G\to P\,\}_{\nu\in{\cal N}}\,$ 
be a bundle atlas. The transition functions, 
$\,\psi_{\nu\mu}\,:\,{\cal U}_\nu\cap {\cal U}_\mu\to G\,$,  
are specified by  
$\,\varphi_\nu(x,g)=\varphi_\nu(x,\psi_{\mu\nu}(x)\cdot g)$. 
The associated local sections, $\,\sigma_\nu:\,{\cal U}_\nu\to P\,$, 
are given by $\,\sigma_\nu(x) = \varphi_\nu(x,\mathbf 1)$. 

Let $\,A\,$ be a connection on $\,P$.  Locally we may express $\,A\,$ 
via $\,A_\nu = \sigma_\nu^* A$. The local connection forms satisfy 
$\,A_\nu\,=\,\psi^{-1}_{\mu\nu}\,A_\mu\,\psi_{\mu\nu}\,
+\,\psi^{-1}_{\mu\nu}\,d\psi_{\mu\nu}$. We say that $\,A\,$ is an 
$\,L^2_{loc}$-connection if all $\,A_\nu\,$ are locally $\,L^2\,$ 
$1$-forms with values in $\,\frak g$. 
The curvature of $\,A\,$ is defined as $\,F_A\,=\,dA\,+\,\frac12\,[A,\,A]$. 
Locally, we have 
$\,(F_A)_\nu\,=\,\sigma_\nu^* F_A\,=\,dA_\nu\,+\,\frac12\,[A_\nu,\,A_\nu]$, 
which can be interpreted as a distribution for an $\,L^2_{loc}$-connection.
A connection $\,A\,$ is flat if $\,F_A=0$.

For smooth flat connections, 
the usual holonomy map is a map from $\,\pi_1(\Omega, x_0)\,$ 
to $\,G$. This map depends on a base point above $\,x_0\,$ in $\,P$. 
Changing the base point in $\,P\,$ will conjugate the holonomy map 
by an element of $G$. The equivalence class of the holonomy map 
up to conjugation is called the holonomy representation. 
We are about to generalize the notion of holonomy representation 
to distributionally flat $\,L^2_{loc}\,$ connections. Note that the value 
of the local gauge, $u$, from the nonlinear Poincar\'e lemma in the 
previous subsection, is not defined at every point. This is why we 
do not generalize the holonomy map. To generalize the holonomy 
representation, we will triangulate the manifold $\,\Omega\,$ and 
use the edge-path group in place of the fundamental group. 

We now review the definition of the edge-path group of a pointed simplicial 
complex, \cite{Spanier}. Let $\,K\,$ be a simplicial complex. 
Denote the vertex set of $\,K\,$ by $\,K^{(0)}$. Select a vertex 
$\,p_0\,$ as the base point. In general, 
vertices 
will be denoted by letters $\,p, q, \dots$, the oriented edges 
will be denoted by $\,e=[p,q]\,$ where $\,i(e)= p\,$ is the 
initial vertex and  $\,f(e)=q\,$ is the final vertex of $e$. 
The closed star of the vertex $p$ 
will be denoted by $\,\hbox{st}(p)$. 
An edge-path, $\,\zeta$, is a non-empty finite sequence of oriented 
edges $\,[e_1, \dots, \,e_\ell]\,$ with $\,f(e_j)=i(e_{j+1})$. 
A closed, pointed edge-path is one with $\,i(e_1)=f(e_\ell)=p_0$. 
Two edge-paths, $\,\zeta_1\,$ and $\,\zeta_2$, are called simply equivalent 
if one can be obtained from the other by replacing a single edge, 
$\,e=[p,r]\,$ by a two-edge path $\,[p,q][q,r]$ 
(or vice versa), where $\,p$, $\,q$, and $\,r$ 
are (not necessarily distinct) vertices of the same $2$-simplex in $\,K$.
Two edge-paths are equivalent if one can be obtained from the other by 
a finite sequence of such moves. The set of equivalence classes of 
all closed pointed edge-paths forms a group, called the edge-path group, 
$\,E(K,\,p_0)$. This group is canonically isomorphic to the 
fundamental group, $\,\pi_1(|K|,\,p_0)$, \cite{Spanier}, Theorem 3.6.17.

It is known that any smooth manifold admits a unique PL-structure 
compatible with the smooth structure, \cite{Munkres}. 
In particular, there exists a simplicial complex, $\,K\,$ whose topological 
realisation, $\,|K|$, is homeomorphic to $\,\Omega$. 
Furthermore, we may assume that any nonempty intersection 
of closed stars of vertices is piecewise smoothly equivalent to the unit cube 
in $\,\mathbb R^n$.  Any two such complexes have a common subdivision. 
Given any open cover of $\,\Omega$, subdividing the complex, if necessary, 
we may assume that the closed star of any vertex is contained in some element 
of the cover. 
Fix a complex $\,K$, retaining the properties 
described above, subordinate to a bundle atlas and fix a vertex, 
$\,p_0$, as the base point. 
Let $\,\phi:\,E(K,\,p_0)\to \pi(|K|,\,p_0)\,$ be the canonical 
isomorphism.   
For each vertex $\,p\,$ choose an index $\,\nu(p)\,$ with 
$\,\hbox{st}(p)\subset {\cal U}_{\nu(p)}$. Denote 
$\,\psi_{\nu(q)\nu(p)}\,$ by $\,\psi_{qp}$, $\,\varphi_{\nu(p)}\,$ by 
$\,\varphi_p$, and $\,A_{\nu(p)}\,$ by $\,A_p$.  

Let us define the holonomy representation of a distributionally 
flat $\,L^2_{loc}\,$ connection, $\,A$.  
By the nonlinear Poincar\'e lemma,  there exists 
$\,u_p\in W^{1,2}(\hbox{st}(p),G)\,$ so that $\,A_p= u_p^{-1}\,du_p$. 
Pick such $\,u_p\,$ for each vertex $\,p$. A collection of such functions is 
called a local developing map. 
We have, 
$$
\begin{array}{rl}
(u_q\,\psi_{qp})^{-1}\,d(u_q\,\psi_{qp}) & 
= \psi_{qp}^{-1}\,u_q^{-1}\,du_q\,\psi_{qp}\,+\,\psi_{qp}^{-1}\,d\psi_{qp}\\
& =\,\psi_{qp}^{-1}\,A_q\,\psi_{qp}\,+\,\psi_{qp}^{-1}\,d\psi_{qp} 
=\,u_p^{-1}\,du_p\,.
\end{array} 
$$
From the second part of the nonlinear Poincar\'e lemma it follows, 
that there exists a $\,g_{[p,q]}\in G\,$ so that 
%\begin{equation}\label{edge}
$$ 
u_p\,=\,g_{[p,q]}\,u_q\,\psi_{qp}\;.
$$
%\end{equation}

%%%%%%%%%

\begin{definition} The holonomy representation of a flat connection, $\,A$,
is the conjugacy class of the map $\,\rho_A:\,\pi_1(\Omega,\,p_0)\,\to\,G\,$ 
given by 
$$
\rho_A(\phi([e_1,\dots,\,\,e_\ell]))\,=\,g_{e_1}\cdot\dots\cdot g_{e_\ell}\;.
$$
\end{definition}

%%%%%%%%%%%%%%

\begin{lemma}
This definition is independent of the representative of the edge-path group, 
the choice of $\,u_p$, $\,\varphi_p$, and the triangulation $\,K$.
\end{lemma} 

%%%%%%%%%%%%%%%%%%

\medskip
\noindent{\bf Proof.} Consider a simple equivalence of 
two edge-paths generated by 
$\,[p,r]\,\leftrightarrow\,[p,q][q,r]$.  Observe that 
$\,\psi_{rq}\psi_{qp}\,=\,\psi_{rp}$. Since  $\,p$, $\,q$, and $\,r$ 
lie within a single simplex, the set
$\,V_{p,q,r}=\hbox{st}(p)\cap\hbox{st}(q)\cap\hbox{st}(r)\,$ is nonempty. 
By the definition of $\,g_e$,   
$$
\begin{array}{c}
g_{[p,r]}\cdot u_r|_{V_{p,q,r}}\cdot \psi_{rp}\, =\,u_p|_{V_{p,q,r}}  
 =\,
g_{[p,q]}\cdot u_q|_{V_{p,q,r}}\cdot \psi_{qp}\\ 
=\,g_{[p,q]}
\cdot g_{[q,r]}\cdot u_r|_{V_{p,q,r}}\cdot \psi_{rq}\,
\cdot \psi_{qp} 
 =\,
g_{[p,q]}\cdot g_{[q,r]}\cdot u_r|_{V_{p,q,r}}\cdot \psi_{rp}
\end{array}
$$
Thus, $\,g_{[p,q]}\cdot g_{[q,r]}\,=\,g_{[p,r]}\,$ and $\,\rho_A\circ \phi\,$ 
respects simple equivalence. 

By the nonlinear Poincare lemma, 
any other choice of local developing map, 
$\,u_p^\prime\,$, is related to $\,u_p\,$ by $\,u_p^\prime = h_p\cdot u_p$,
for some constants $\,h_p$. The corresponding edge labels are 
given by $\,g^\prime_{[p,q]}=\,h_p\cdot g_{[p,q]}\cdot h_q^{-1}$. 
This implies that $\,\rho_A\,$ changes by conjugation by $\,h_{p_0}$, i.e., 
$\,\rho^\prime_A(\gamma)= h_{p_0}\cdot \rho_A(\gamma)\cdot h_{p_0}^{-1}$. 

Let $\,K^\prime\,$ be a subdivision of $\,K$, and 
let $\,\phi^\prime:\,E(K^\prime,p_0)\to\pi_1(|K^\prime|=|K|,p_0)\,$ 
be the corresponding isomorphism. Any loop in $\,\Omega\,$ is represented by 
an edge-path 
$\,\zeta=[e_1,\dots,e_\ell]\,$ in $\,K$. In $\,K^\prime\,$ the same loop 
is represented by 
a subdivision  
$\,\zeta^\prime=[e_{1,1},\dots,e_{1,k_1} ,\dots,e_{\ell,k_\ell}]\,$. 
An edge, $\,[p,q]$, of $\,\zeta\,$ is subdivided into a subpath 
$\,[[p=r_1,r_2],\dots,[r_{k-1},r_k=q]]$. The complex $\,K^\prime\,$ 
is subordinate to the same bundle atlas as $\,K$. Thus, we may choose 
$\,\nu(r_1)=\dots =\nu(r_{m})=\nu(p)\,$ and 
$\,\nu(r_{m+1})=\dots =\nu(r_{k})=\nu(q)$, and also, 
$\,u_{r_1}=\dots =u_{r_{m}}=u_p\,$,  
$\,u_{r_{m+1}}=\dots = u_{r_{k}}= u_{q}$, and 
$\,\phi_{r_1}=\dots = \phi_{r_{m}}=\phi_p\,$,  
$\,\phi_{r_{m+1}}=\dots = \phi_{r_{k}}= \phi_{q}$. 
This implies  
$\,g^\prime_{[p=r_1,r_2]}\cdot\dots\cdot g^\prime_{[r_{k-1},r_k=q]}
=\,{\mathbf 1}\cdot\dots \cdot {\mathbf 1}\cdot 
g_{[p,q]}\cdot {\mathbf 1}\cdot\dots \cdot {\mathbf 1}$. 
Thus, $\,\rho_A\,$ does not depend on the triangulation. 

Let $\,\{\varphi^\prime_\nu :\,{\cal U}_\nu\times G\to P\}\,$ be 
a second bundle atlas. By taking intersections, if necessary, 
we assume 
that it has the same open cover $\,\{{\cal U}_\nu\}$. Subdividing, 
if necessary, we may assume that the complex $\,K\,$ is the same. 
We may therefore pick the same indexing functions, $\,\nu(p)$.  
There exist transition functions $\,\vartheta_\nu:\,{\cal U}_\nu\to G\,$ 
such that 
$\,\varphi^\prime_\nu(x,g)\,=\,\varphi_\nu(x,\vartheta_\nu(x)\, g)$.
We will put primes on all objects computed using the second bundle chart. 
The transition functions satisfy 
$\,\psi^\prime_{\mu\nu}\,=\,\vartheta_\mu^{-1}\,\psi_{\mu\nu}\,\vartheta_\nu$. 
The local connection forms satisfy 
$$
\begin{array}{rl}
A^\prime_p\,&=\, \vartheta_p^{-1}\,A_p\,\vartheta_p\,+\,
\vartheta_p^{-1}\,d\vartheta_p \\ 
&=\,\vartheta_p^{-1}\,u_p^{-1}\,du_p\,\vartheta_p\,+\,
\vartheta_p^{-1}\,u_p^{-1}\,u_p\,d\vartheta_p \\ 
&=\,(u_p\,\vartheta_p)^{-1}\,d(u_p\,\vartheta_p)\,, 
\end{array}
$$
and we have $\,u^\prime_p\,=\,u_p\,\vartheta_p$. 
Also, $\,g^\prime_{[p,q]}\,=\,g_{[p,q]}$. Indeed, 
$$
g^\prime_{[p,q]}\,=\,u^\prime_p\,\psi^\prime_{pq}\,(u^\prime_q)^{-1}\,
=\,u_p \vartheta_p\,\psi^\prime_{pq}\,\vartheta_q^{-1} u_q^{-1}\,=\,
u_p \psi_{pq}\, u_q^{-1}\,=\,g_{[p,q]}.
$$
This completes the proof of the lemma.

%%%%%%%%%%%%%

\begin{remark} One can prove that for smooth connections our definition 
of the holonomy representation agrees with the usual one.
\end{remark}

%%%%%%%%%%%%%%%%%%

\section{Configuration spaces and the Skyrme functional}\label{sec:4}

In this section we generalize the original Skyrme model to two 
separate settings. In the first setting the fields are maps from 
a $3$-manifold to a compact Lie group. In the second setting, 
the fields are flat connections on a principle bundle over a $3$-manifold.  
 
To define the Skyrme functional for maps, fix a Riemannian metric 
on the base manifold $\,M$, and fix a faithful unitary representation 
of the compact Lie group $\,G$. Define the norm on the Lie algebra $\,\frak g\,$ 
as $\,|X|^2\,=\,-\frac18\,\hbox{Tr}(\hbox{ad}(X)\hbox{ad}(X))$. 
With the help of the Riemannian metric, this extends to 
a norm on Lie algebra valued forms in a standard way. 
The Skyrme functional for maps is defined to be
$$
E(u)=\int_{M} \frac12 |u^{-1}\,d u|^2\,
+\,\frac14 |u^{-1}\,d u\wedge u^{-1}\,d u|^2\;d\,\hbox{vol}_M\,.
$$
The configuration space of Skyrme maps is defined to be 
$$
{\cal S}^G(M)\,=\,\{u\in W^{1,2}(M,G)\,|\,E(u)<\infty\}/\sim\,,
$$
where the equivalence is given by $\,u(\cdot)\;{\sim}\; u(\cdot)\, g$, 
for $\,g\in G$.

To define the Skyrme functional for connections, we need a special 
type of reference connection. We start with the description of the 
admissible reference connections. Let $\,P\,$ be a principal $\,G$-bundle 
over $\,M$, 
and let $\,B\,$ be a (not necessarily flat) smooth connection on $\,P$. 
Given a base point $\,x_0\in M\,$ and a lift $\,\hat x_0\in P\,$ of $\,x_0$, 
we define a holonomy homomorphism 
$\,\varrho_B :\,\Omega_{x_0}\to G$, 
where $\,\Omega_{x_0}\,$ is the based loop group 
of $\,M$, as follows. Given a based loop $\,\gamma\,$ in $\,M$, 
construct the horizontal lift $\,\tilde\gamma:\,[0,1]\to P\,$ 
with $\,\tilde\gamma(0)=\hat x_0$. By definition,  
$\,\varrho_B(\gamma)\,$ is the unique element of $\,G\,$ satisfying 
$\,\tilde \gamma(1)\cdot \varrho_B(\gamma)\,=\,\hat x_0$. 
Note that the horizontal lift $\,\tilde\gamma\,$ 
may be constructed from any lift $\,\hat\gamma\,$ starting at $\,\hat x_0\,$ 
as $\,\tilde\gamma(t)\,=\,\hat\gamma(t)\,g(t)$, where  
$\,g:\,[0,1]\to G\,$ is the solution of the equation 
$\,\dot g(t) + i_{\dot{\tilde\gamma}(t)}\,B\,g(t)\,=\,0\,$ 
with $\,g(0)=\mathbf 1$. 

%%%%%%%%%%%%%%%%

\begin{remark} In this paper we have used three flavors of 
holonomy: the holonomy homomorphism, the holonomy map, and the holonomy 
representation. The holonomy homomorphism is defined for arbitrary smooth 
connections $\,B$. 
If the connection $\,B\,$ is flat, then the holonomy homomorphism 
$\,\varrho_B\,$  
depends only on the homotopy class of $\,\gamma$, and therefore 
reduces to the holonomy map, $\,\rho_B:\,\pi_1(M, x_0)\to G$. 
If the lift of the base point $\,\hat x_0\,$ is changed, then 
the holonomy map changes by conjugation. The equivalence class 
of $\,\rho_B\,$ up to conjugacy is the holonomy representation. 
\end{remark}

%%%%%%%%%%%%%%

A connection $\,B\,$ is called central if $\,\varrho_B(\Omega_{x_0})\,$ 
is contained in the center, $\,Z_G$, of $\,G$. A principal bundle is 
called central if it admits a central connection. Clearly, 
any bundle with abelian structure group is central. In general, 
a principal $\,G$-bundle $\,P\,$ is central if and only if 
$\,P\,\times_G\,(G/Z_G)\,$ is a trivial bundle. 
For connected Lie groups, there is 
an obstruction in $\,H^2(M;\widetilde{\pi_1(G/Z_G)})\,$ which vanishes 
if and only if
the bundle $\,P\,\times_G\,(G/Z_G)\,$ is trivial.  

Let $\,P\,$ be a central bundle with central connection $\,B$. 
Any other connection on $\,P\,$ may be expressed as 
$\,A = B\,+\,\hbox{pr}^* a$, where $\,a\,$ is a $\,\frak g$-valued $1$-form 
on $\,M$. Our generalization of the Skyrme functional 
uses this description:
$$
E[a]=\int_{M} \frac12 |a|^2\,+\,\frac1{16} |[a,\,a]|^2\;d\,\hbox{vol}_M\,.
$$  
The corresponding configuration space of Skyrme potentials is defined to be 
$$
{\cal S}^{B,G}[M]\,=\,\{ a\in L^2(M,\frak g)\,|\,
da + \frac12\,[a,\,a]\,+\,F_B\,=\,0,\; E[a]<\infty\}/\sim\;,
$$
where $\,a\,\sim\, g^{-1}\,a\,g\,$ for $\,g\in G$.

%%%%%%%%%%%%%

\begin{remark} The Skyrme functional is not gauge invariant. 
It is only invariant 
under constant gauge transformations. This is why the configuration space 
of Skyrme fields is infinite dimensional.
\end{remark}

%%%%%%%%%%%%%%%

The transformation $\,u\,\mapsto u^{-1} du\,$ takes the configuration 
space of Skyrme maps into the configuration space of Skyrme potentials 
with the trivial reference connection. We may view this 
as a transformation 
$\,{\cal D}:\,{\cal S}^G(M)\to {\cal S}^{\theta, G}[M]$, 
or as a transformation 
$\,\tilde{\cal D}:\,{\cal S}^G(M)\to {\cal S}^{\tilde\theta,\tilde G}[M]$, where 
$\,\theta\,$ ($\,\tilde\theta$) 
is the trivial connection on $\,M\times G\,$ ($\,M\times\tilde G$). We have 
$\,E(u)=E[{\cal D}u]= E[\tilde{\cal D}u]$. We will reconsider the map 
$\,{\cal D}\,$ at the end of this section.

%%%%%%%%%%%%%%%%

In Section \ref{sec:2} we described the path components of 
the space of smooth maps from $\,M\,$ to $\,G\,$ up to 
multiplication by $\,G\,$ on the right. It turns out that the 
numerical invariants specifying these components, 
Proposition 2,  are well defined for 
Sobolev maps.  To show this, we combine the description 
of these invariants given before Proposition 2 with the general 
definition of the holonomy representation from Section \ref{sec:3}.
We need to relate the holonomy of a connection on the trivial bundle 
$\,M\times G\,$ to the holonomy of the corresponding connection on 
the trivial bundle $\,M\times\tilde G$. The following lemma 
addresses a slightly more general situation in which 
the standard map  $\,\tilde G\to G\,$ 
is replaced by a Lie group homomorphism $\,H\to G$. 

%%%%%%%%%%%%%%

\begin{lemma} Let $\,f:\,H\to G\,$ be a  homomorphism 
of Lie groups such that $\,f_*:\frak h\to \frak g\,$ is an isomorphism. 
Let $\,\hat f:\,M\times H\to M\times G\,$ be the obvious map. 
If $\,A\,$ is a flat connection on $\,M\times G$, then 
$\,\tilde A\,=\,f^{-1}_*\,\hat f^*\,A\,$ 
is a flat connection on $\,M\times H$, and 
the holonomy representations are related by the formula: 
$\,\rho_A\,=\,f\circ \rho_{\tilde A}$.
\end{lemma}

\noindent{\bf Proof.\/} Let $\,R_k\,$  denote right multiplication 
by $\,k\,$ in a given Lie group. Let $\,L_s\,$ denote left multiplication 
by an element $\,s\,$ of any space with a right group action. 
First note that $\,\tilde A\,$ 
is right invariant. 
Indeed, 
$$
R_h^*\tilde A = f^{-1}_*\,R_h^*\,\hat f^* A 
= f^{-1}_*\,\hat f^*\,R_{f(h)}^*\,A = 
f^{-1}_*\,\hat f^*\,(\hbox{Ad}\,f(h)^{-1})_*\,A = 
(\hbox{Ad}\,h^{-1})_* \tilde A\,,
$$
where we have used the equalities $\,\hat f\circ R_h = R_{f(h)}\circ \hat f\,$ 
and $\,(\hbox{Ad}\,f(h))\circ f\,=\,f\circ (\hbox{Ad}\,h)$.  
Now notice that $\,\tilde A\,$ satisfies the second condition in the 
definition of a connection. Indeed, 
$$
\tilde A (L_{(x,h)\,*}X)=f^{-1}_*\,A (\hat f_*L_{(x,h)\,*}X)=
f^{-1}_*\,A (L_{(x,f(h))\,*}\,f_*X)=f^{-1}_*\,f_*\,X=X\,,
$$ 
for any $\,X\,$ in $\,\frak h$, since 
$\,\hat f\circ L_{(x,u)}\,=\,L_{(x,f(h))}\circ f$. 
The connection $\,\tilde A\,$ is flat because $\,A\,$ is. 

To compute the holonomy representations we use the trivial bundle atlases  
and a fixed triangulation. Let $\,p\,$ be a vertex of the 
triangulation and $\,\tilde \sigma_p\,$ denote the local section 
$\,\tilde \sigma_p :\,\hbox{st}(p)\to M\times H\,$ which is simply 
$\,\tilde \sigma_p(x)=(x,\mathbf 1)$. Then, 
$\,\sigma_p\,=\,\hat f\circ \tilde\sigma_p$. Let $\,\tilde u_p\,$ 
be the associated local developing map for $\,\tilde A$. 
The following computation shows that  
$\,u_p\,=\,f\circ\tilde u_p\,$ is a local developing map for $\,A$. 
$$
\begin{array}{c}
u_p^{-1}du_p = f(\tilde u_p^{-1})\,f_* d\tilde u_p 
= f(\tilde u_p^{-1})\,f_* \tilde u_p\,\tilde u_p^{-1}d\tilde u_p  
=  f(\tilde u_p^{-1})\,f_* (L_{\tilde u_p})_*\,\tilde \sigma^*_p\,
f^{-1}_*\hat f^*\,A \\ 
= 
f(\tilde u_p^{-1})\,f_* (L_{\tilde u_p})_*\,f^{-1}_* \sigma^*_p\,A 
= f(\tilde u_p^{-1})\,(L_{f(\tilde u_p)})_*f_*\,f^{-1}_* \sigma^*_p\,A 
= \sigma^*_p\,A\,.
\end{array}
$$
Here we used $\,f\circ L_{\tilde u_p}\,=\,L_{f(\tilde u_p)}\circ f$. 
It follows that $\,g_e=f(\tilde g_e)$, and so,  
$\,\rho_{\tilde A}=f\circ \rho_A\,$ establishing the lemma.

%%%%%%%%%%%%%%%%%%%%%%%%%

Returning to the construction of  invariants for Sobolev maps, 
we will apply the previous lemma to the standard map 
$\,\tau:\,\tilde G\to G_0\to G$. When $\,u\,$ is an element of 
$\,{\cal S}^G(M)$, the form $\,A\,=\,\theta +\hbox{pr}^*{\cal D}u \,$
is a connection on $\,M\times G$. Here $\,\theta\,$ is the trivial connection 
on $\,M\times G$. Notice that $\,u\,$ is a local developing map for $\,A\,$ 
associated to the trivial bundle atlas. The corresponding holonomy 
representation, $\,\rho_A$, is therefore trivial. 
Now, for $\,\tilde A\,=\,\tilde \theta +\hbox{pr}^*\tilde{\cal D}u = 
\tau^{-1}_*\hat\tau^* A$, the previous lemma implies 
$\,\rho_{\tilde A} = \tau\circ\rho_A\,=\,\mathbf 1$. Since 
$\,\tau^{-1}(\mathbf 1)\cong \pi_1(G_0)\cong H_1(G_0)$, we have 
$\,\rho_{\tilde A}(\pi_1(M))\subset Z_{\tilde G}$. 
In general, the holonomy is only well defined up to conjugation, 
but in this case, the holonomy is a well-defined map, 
$\,\rho_{\tilde A}:\,\pi_1(M)\to H_1(G_0)$. Since $\,H_1(G_0)$ is abelian, 
this must factor through a map from $\,H_1(M)\,$ to $\,H_1(G_0)$. 
Denote the corresponding element of $\,H^1(M; H_1(G_0))\,$ by $\,\alpha_u$. 
The constants $\,a_\ell(u)\,$ from Proposition 2 are just the coordinates of 
$\,\alpha_u\,$ in the basis $\,\{\alpha_\ell\}$ of $\,H^1(M; H_1(G_0))$. 

Given any element $\,\alpha\in H^1(M; H_1(G_0))$, there exists 
a smooth map $\,v_\alpha:\,M\to G_0\,$ with $\,(v_\alpha)_*=\alpha$. 
Fix such a map for each cohomology class. Combining the above discussion 
with  Remark \ref{valpha} and Remark \ref{r2}, we see that the integral 
$\,\int_M (u\,(v_{\alpha_u})^{-1})^*\Theta^k\,$ 
is well defined for Sobolev maps.  
We partition the configuration space of maps into the following sectors:
$$
{\cal S}^G_{\alpha; c^1,\dots,c^N}(M)\,=\,
\{u\in{\cal S}^G(M)\,|\,\alpha_u=\alpha,\; 
\int_M (u\,(v_{\alpha_u})^{-1})^*\Theta^k\,=c^k\}\,.
$$

We do not yet know that two maps have  
the same invariants if and only if they 
live in the same path component. It is true for smooth 
maps, and it would be true if we knew that any finite Skyrme energy 
Sobolev map could be approximated
by smooth maps with uniformly bounded Skyrme energy.

We now turn to Skyrme potentials. As we have shown, 
the holonomy is well defined for $\,L^2_{loc}\,$ distributionally 
flat connections. We first partition $\,{\cal S}^{B,G}[M]\,$ 
using the holonomy representation as follows:
$$
{\cal S}^{B,G}_{\rho}[M]\,=\,
\{a\in{\cal S}^{B,G}[M]\,|\,\rho_{B+a}=\rho\}\,.
$$
In the smooth case it is well known that two flat connections are 
gauge equivalent if and only if they have the same holonomy. 
To describe the path components of the space of 
smooth flat connections with fixed holonomy, it is convenient to 
fix a reference connection in this class. Let $\,B+b\,$ be one 
such connection. Any other connection, $\,B+a$, is obtained from
$\,B+b\,$ by a gauge transformation, 
$\,B+a = (B+b)^u = B+u^{-1}\,b\,u + u^{-1} d_B u$, where 
$\,u\,$ is a map from $\,M\,$ to $\,G$. 
Conversely, given any such $\,u$, one obtains 
a connection $\,(B+b)^u\,$ with the same holonomy. 
The connections with fixed holonomy modulo constant gauge 
transformation are in one-to-one correspondence with 
the maps $\,u:\,M\to G\,$ modulo right multiplication by elements of $\,G$. 
The path components of the space of equivalence classes 
of smooth flat connections with fixed holonomy 
are determined by the one dimensional invariant and the three dimensional 
invariant of the gauge, $\,u$, as described above. 
We will now show that these invariants are well defined for 
Skyrme potentials with fixed holonomy.

We start by showing that two  $\,L^2_{loc}\,$ distributionally 
flat connections have the same holonomy if and only if 
they are gauge equivalent. We only prove this for connections on 
central bundles. 

Let $\,\Omega\,$ be a Riemannian manifold with sufficiently 
smooth boundary. For central principal $\,G$-bundles over $\,\Omega\,$ 
we define the gauge group to be $\,{\cal G}\,=\,W^{1,2}_{loc}(\Omega, G)$, 
see Remark \ref{remark3}.

%%%%%%%%%%%%%%%%%%%%%%%%

\begin{lemma}\label{gaugeflat}
Two $\,L^2_{loc}\,$ distributionally 
flat connections on a central bundle are gauge equivalent if 
and only if they have the same holonomy.
\end{lemma}

\noindent{\bf Proof.\/} Let $\,B\,$ be a smooth 
central reference connection. Let $\,A = B + a\,$ 
be an $\,L^2_{loc}\,$ distributionally 
flat connection, and  take $\,g\in{\cal G}$. The element $\,g\,$ acts 
on $\,A\,$ by $\,A^g = B + g^{-1} a\,g + g^{-1} d_B g$. 
To compute the holonomy representation of each connection, 
we fix a central bundle atlas and a triangulation. 
On the star of the vertex $\,p\,$ 
we have $\,A_p\,=\,B_p + a_p\,$ and 
$\,A^g_p\,=\,B_p + g^{-1}a_p\,g + g^{-1}dg$. Pick a developing map $\,u_p\,$ 
for $\,A_p$, i.e., $\,u^{-1}_p du_p = A_p$. Then $\,u^g_p =  u_p\,g\,$ 
will be a developing map for $\,A^g_p$. Now, $\,u_p = g_{[p,q]}u_q\,\psi_{qp}$, 
so $\,u_p\cdot g = g_{[p,q]}u_q\,\psi_{qp}\cdot g =g_{[p,q]}u_q\,g\,\psi_{qp}$. 
Thus, the holonomy representation is the same. 

Now, assume that $\,A^1=B+a^1\,$ and $\,A^2=B+a^2\,$ have the same holonomy 
representation. We fix a central bundle atlas and a triangulation. 
Pick local developing maps $\,u^1_p\,$ and $\,u^2_p$, and let $\,g^1_{[p,q]}\,$
and $\,g^2_{[p,q]}\,$ denote the corresponding edge labels. 
The holonomy representation is defined up to conjugacy. 
Multiplying $\,u^2\,$ at the base vertex $\,p_0\,$ by a constant 
we may assure that the holonomy maps of $\,A^1\,$ and $\,A^2\,$ are the same. 
Choose a maximal tree in the $1$-skeleton of the triangulation with root 
$\,p_0$. Any vertex is connected to the root by a unique path, 
$\,(p_0,p_1,\dots, p_m)$. We inductively modify the local developing map 
$\,u^2\,$ as follows. 
Set $\,\bar u^2_{p_0}= u^2_{p_0}$, and 
recursively define 
$$
\bar u^2_{p_{i+1}}\,=\,g^1_{[p_{i+1},p_i]}\,g^2_{[p_{i},p_{i+1}]}\, 
u^2_{p_{i+1}}\,.
$$  
The required gauge transformation is defined on each star by 
$\,g\,=\,(u^1_p)^{-1}\,\bar u^2_p$. On the overlaps corresponding to 
the edges of the tree, these functions agree by construction. 
Any edge, $\,e$, not in the maximal tree belongs to a circuit, say, 
$\,[[q_1,q_2],\dots,[q_{m-1}, q_m], [q_m, q_1]]$, with $\,q_1=p_0\,$ and 
each edge $\,[q_i,q_{i+1}]$, except $\,e$, in the tree. 
Let $\,\bar g^2_f\,$ be the edge labels constructed from $\,\bar u^2$. 
When $\,f\,$ is in the tree, $\,\bar g^2_f = g^1_f\,$ by construction.  
The product $\,g^1_{[q_1,q_2]}\dots g^1_{[q_{m-1}, q_m]} g^1_{[q_m, q_1]}\,$ 
is equal to 
$\,\bar g^2_{[q_1,q_2]}\dots \bar g^2_{[q_{m-1}, q_m]} \bar g^2_{[q_m, q_1]}$, 
because the holonomies agree. Hence, $\,\bar g^2_e = g^1_e$, and so 
$\,\bar g^2_f = g^1_f\,$ for every edge $\,f$. Now, 
$\,(u^1_p)^{-1}\bar u^2_p 
= (u^1_q)^{-1}g^1_{[q,p]}\bar g^2_{[p,q]}\bar u^2_q = (u^1_q)^{-1}\bar u^2_q$,
for any two neighboring vertices $\,p\,$ and $\,q$. This 
implies that the function $\,g\,=\,(u^1)^{-1}\bar u^2\,$ 
is well defined globally on $\,\Omega$. It belongs to 
$\,W^{1,2}_{loc}(\Omega,G)$, 
and it is easy to check that $\,(A^1)^g \,=\,A^2$. The lemma is proved.

%%%%%%%%%%%%%%

Returning to Skyrme potentials on a closed $3$-manifold $\,M$, 
fix a reference connection $\,B\,$ and holonomy $\,\rho$. 
Fix $\,b\in{\cal S}^{B,G}_\rho\,$ as a reference field. 
Given any other field $\,a\in{\cal S}^{B,G}_\rho[M]$, by 
the previous lemma, there exists  
$\,u\in W^{1,2}(M,G)\,$ such that $\,B+a = (B+b)^u$. The one dimensional 
invariant of $\,a$ is nothing more then $\,\alpha_u\,$ described 
previously. In this setting we will denote it $\,\alpha[a]$. 
Define $\,{\underline u}\,$ to be $\,{\underline u} = u v^{-1}_{\alpha[a]}$. 

The three dimensional invariants are given by 
$$
c^k[a]\,=\,-\frac{K_{\tilde G^k}}{192\,\pi^2}
\int_M \hbox{Tr}(\hbox{ad}([\widehat{{\underline u}^{-1}d{\underline u}},\,
\widehat{{\underline u}^{-1}d{\underline u}}])\,
\hbox{ad}(\widehat{{\underline u}^{-1}d{\underline u}}))\,.
$$
Here $\,k\,$ is the index corresponding to the simple 
factor $\,\tilde G^k\,$ of the covering group $\,\tilde G$,  
and the constants $\,K_{\tilde G^k}\,$ are those presented in 
Table \ref{tab:1}, see Section \ref{sec:2.2}. The $\,\widehat\,$\ over 
$\,{\underline u}^{-1}d{\underline u}\,$ means orthogonal projection 
onto the Lie 
algebra of $\,G^k$. We include $\,v^{-1}_\alpha\,$ to ensure 
that the invariants are integral in the smooth case.

%%%%%%%%%

\begin{lemma}\label{3inv} 
The invariant $\,c^k[a]\,$ is well defined for 
Skyrme potentials with fixed holonomy.
\end{lemma}
\noindent{\bf Proof.\/} Given a reference potential 
$\,b\in {\cal S}^{B,G}_\rho$, any other potential, $\,a$, with the same holonomy 
satisfies $\,a\,=\,u^{-1} b u + u^{-1}du$, for some $\,u\in W^{1,2}(M,G)$. 
Since
$$
|\hbox{Tr}(\hbox{ad}([\widehat{{\underline u}^{-1}d{\underline u}},\,
\widehat{{\underline u}^{-1}d{\underline u}}])\,
\hbox{ad}(\widehat{{\underline u}^{-1}d{\underline u}}))|\,\le\,c\,
|\hbox{Tr}({\underline u}^{-1}d{\underline u}
\wedge {\underline u}^{-1}d{\underline u}
\wedge {\underline u}^{-1}d{\underline u})|\,,
$$
for some constant independent of $\,u$, we will show that 
$\,\hbox{Tr}({\underline u}^{-1}d{\underline u}
\wedge {\underline u}^{-1}d{\underline u}
\wedge {\underline u}^{-1}d{\underline u})\,$ is in $\,L^1$. 
Using $\,u^{-1}du\,=\,a\,-\,u^{-1} b u $, we compute

%%%%%%

\begin{equation}\label{wedge}
\begin{array}{c} 
\hbox{Tr}(u^{-1}du\wedge u^{-1}du\wedge u^{-1}du) = 
\hbox{Tr}(a\wedge a\wedge a - a\wedge a\wedge u^{-1}b u \\
- a\wedge u^{-1}b u\wedge a - u^{-1}b u\wedge a\wedge a 
+ u^{-1}b u\wedge u^{-1}b u\wedge a  \\ + u^{-1}b u\wedge a\wedge u^{-1}b u 
+ a\wedge u^{-1}b u\wedge u^{-1}b u 
- u^{-1}b u\wedge u^{-1}b u \wedge u^{-1}b u )\,.
\end{array}
\end{equation}
When $\,\omega_2\,$ is a matrix-valued $2$-form and $\,\omega_1\,$ is 
a matrix-valued $1$-form, 
$\,\hbox{Tr}(\omega_1\wedge\omega_2)= \hbox{Tr}(\omega_2\wedge\omega_1)$. 
Recall that $\,a,\,b,\,a\wedge a$, and $\,b\wedge b\,$ are all in $\,L^2$, 
and $\,u,\,u^{-1}\in L^\infty$. It follows that each term in (\ref{wedge}) 
is in $\,L^1$. Observe, that 
$\,{\underline u}^{-1} d{\underline u} = 
v_{\alpha[a]}(u^{-1} du- v^{-1}_{\alpha[a]} dv_{\alpha[a]})v_{\alpha[a]}^{-1}$.  
Hence,

%%%%%%%%%%

$$
\begin{array}{c} 
\hbox{Tr}({\underline u}^{-1}d{\underline u}
\wedge {\underline u}^{-1}d{\underline u}
\wedge {\underline u}^{-1}d{\underline u}) = 
\hbox{Tr}(u^{-1}du\wedge u^{-1}du\wedge u^{-1}du \\ 
 - 3\,u^{-1}du\wedge u^{-1}du\wedge v^{-1}_{\alpha[a]} dv_{\alpha[a]} \\
+3\, u^{-1}du\wedge v^{-1}_{\alpha[a]} dv_{\alpha[a]}\wedge 
v^{-1}_{\alpha[a]} dv_{\alpha[a]}   \\ 
- v^{-1}_{\alpha[a]} dv_{\alpha[a]}\wedge v^{-1}_{\alpha[a]} dv_{\alpha[a]}
\wedge v^{-1}_{\alpha[a]} dv_{\alpha[a]})\,.
\end{array}
$$

By the previous computation, the first term is in $\,L^1$. 
Each of the remaining terms is in $\,L^1\,$ because $\,u^{-1}du\in L^2\,$ 
$\,v_{\alpha[a]}\,$ is smooth.  
This proves the lemma. 

\medskip
 
%%%%%%%%%%%%%%%%%%%%%%%%%%
 
Partition the space of Skyrme potentials with fixed holonomy into
the following sectors:
$$
{\cal S}^{B,b,G}_{\rho, \alpha; c^1,\dots,c^N}[M]\,=\,
\{a\in {\cal S}^{B,G}_{\rho}[M]\,|\,\alpha[a]=\alpha, \,c^k[a]=c^k\}\,.
$$
As in the case of Skyrme maps, we will show that any two $\,L^2\,$ Skyrme 
potentials from the same path component are in the same sector. 
If we knew that every finite energy Skyrme potential could be approximated 
by smooth Skyrme potentials with bounded energy, then 
we would conclude that the sectors are the path components.

%%%%%%%%%%%%%%%%%%%%%%%

We conclude this section with a remark about $\,{\cal D}$. 

%%%%%%%%%%%%%%%%

\begin{remark}\label{D} The map 
 $$
{\cal D} :\, {\cal S}^{G}_{ \alpha; c^1,\dots,c^N}(M)\,\to 
{\cal S}^{\theta,0,G}_{{\mathbf 1}, \alpha; c^1,\dots,c^N}[M]
$$
is well defined. Indeed, by the very definition $\,{\cal D}\,$ 
preserves the invariants.  
To show that 
$\,{\cal D} u\,$ is ditributionally flat 
for any $\,u\in W^{1,2}(M, G)$, consider in a chart the approximating sequence 
$\,(u^{-1})_\epsilon d(u)_\epsilon $, where $\,(\cdot)_\epsilon\,$ 
is the usual mollification. This sequence converges to $\,u^{-1}du\,$ 
in the sense of distributions. The differential, 
$\,d((u^{-1})_\epsilon d(u)_\epsilon)\,
=\,(-u^{-1}du\,u^{-1})_\epsilon\wedge (du)_\epsilon$, converges to 
$\,-u^{-1}du\wedge u^{-1}du\,$ in $\,L^1$, hence, in 
the sense of distributions. 
\end{remark}

%%%%%%%%%%%%%%%%%%%%%%%%

\section{Existence of minimizers}\label{sec:5}

In this section we prove that each sector of the configuration space of 
Skyrme maps containes a minimizer and 
each sector of the configuration space of 
Skyrme potentials containes a minimizer as well. 
We use the direct method of the calculus of variations, i.e., 
we prove that for suitable minimizing sequences the topological 
invariants are preserved and the energy is lower semicontinuous.
We begin by analyzing the holonomy. The following lemma  
establishes convergence of the holonomy and, therefore, one-dimensional 
invariants.

%%%%%%%%%%%%%%%%%%

Let $\,\Omega\,$ be an $\,n$-dimensional Riemannian manifold. 
We define a topology on the space of representations 
$\,\pi_1(\Omega, x_0)\to G\,$ as follows. We say that 
$\,\rho_n\,$ converges to $\,\rho$, if for every finite subset 
$\,\{\gamma_1,\dots, \gamma_k\}\subset\pi_1(\Omega, x_0)\,$ 
there exist representatives $\,\bar\rho_n$, $\,\bar\rho$ so that 
$\,\bar\rho_n(\gamma_j)\to\bar\rho(\gamma_j)\,$ in $\,G$ for every $j$. 
It is not hard to see that the topology defined by this convergence 
is Hausdorff when $\,\pi_1(\Omega, x_0)\,$ is finitely generated.  
We next prove a simple lemma about the holonomy representation. 

%%%%%%%%%%%%%%%%%%%%%

\begin{lemma}\label{rhoweakconv} 
If $A^{(n)}$ is a sequence of $L^2_{loc}$ distributionaly flat connections
that converges weakly to a distributionally flat connection, $A$, then 
there exists a subsequence $A^{(n_k)}$ so that 
$\rho_{A^{(n_k)}}\to \rho_A$. 
\end{lemma}

\noindent {\bf Proof.}  
We are in the setting from Section \ref{sec:3}. 
Let $\,\psi_{qp}\,$ be transition functions for the underlying bundle. 
As in the definition of holonomy, there exist 
functions $\,u_p^{(n)}\,$ and group elements $\,g_{[p,q]}^{(n)}\,$ so that 
$\,A_p^{(n)}\,=\,(u_p^{(n)})^{-1}du_p^{(n)}\,$ and 
$\,u_p^{(n)}\,=\,g_{[p,q]}^{(n)}u_q^{(n)}\psi_{qp}$. 
Since $\,A_p^{(n)}\,$ are uniformly bounded in $\,L^2(\hbox{st}(p))$, 
after taking a subsequence we may assume that $\,u_p^{(n)}\,$ converges 
weakly in $\,W^{1,2}$,  strongly in $\,L^2$, and almost everywhere 
to some $\,u_p$. At the same time, since $\,G\,$ is compact, 
we may also assume that 
$\,g_{[p,q]}^{(n)}\,$ converges to some $\,g_{[p,q]}$. It follows that 
$\,A_p\,=\,(u_p)^{-1}du_p\,$ and 
$\,u_p\,=\,g_{[p,q]}u_q\psi_{qp}$. Since we only need to check 
convergence on a finite set of group elements, this concludes the proof. 

%%%%%%%%%%%%%%%%%%%%%

In order to address the three-dimensional invariants, we will need a 
special case of Tartar's $\,div-curl\,$ lemma,
see \cite{Temple}. 
This lemma will also be used later in the proof of the two main theorems. 

%%%%%%%%%%%%%

\begin{lemma}\label{div-curl} 
Let $\,M\,$ be a smooth $\,3$-dimensional Riemannian manifold. 
Let $\,\omega_1^m\in L^2\,$ be a sequence of matrix-valued $1$-forms 
and $\,\omega_2^m\in L^2\,$ be a sequence of matrix-valued $2$-forms on $\,M$. 
If $\,\omega_1^m\,$ converges weakly in $\,L^2\,$ to a form $\,\omega_1$ 
and $\,\omega_2^m\,$ converges weakly in $\,L^2\,$ to a form $\,\omega_2$, 
and if each sequence $\,d\omega_1^m\,$ and $\,d\omega_2^m\,$ 
is precompact in $\,W^{-1,2}_{loc}(M)$, then 
$\,\omega_1^m\wedge \omega_2^m\,$ converges to $\,\omega_1\wedge \omega_2\,$ 
in the sense of distributions.
\end{lemma}

Returning to the settings of Lemma \ref{3inv}, work
on a closed $3$-manifold $\,M\,$ with a  
fixed a reference field $\,b\in{\cal S}^{B,G}_\rho$. 

%%%%%%%%%%%%

\begin{lemma}\label{cconv} 
Given a sequence $\,a_n\in {\cal S}^{B,b,G}_{\rho,\alpha} [M]\,$ 
and $\,a\in{\cal S}^{B,b,G}_{\rho, \alpha}[M]\,$ 
such that $\,a_n\rightharpoonup a\,$ and 
$\,a_n\wedge a_n\rightharpoonup a\wedge a\,$ in $\,L^2(M)$, 
there exists a subsequence such that $\,c^k[a_n]\to c^k[a]$.
\end{lemma}
\noindent{\bf Proof.\/} 
Given any  field $\,a_n\in{\cal S}^{B,b,G}_{\rho, \alpha}[M]$, by 
Lemma \ref{gaugeflat}, there exists a gauge transformation  
$\,u_n\in W^{1,2}(M,G)\,$ such that 
$\,a_n\,=\,u_n^{-1} b u_n\,+\,u_n^{-1}du_n$. 
Recall that $\,\|u_n\|_{L^\infty}\,$ is uniformly bounded, therefore 
$\,du_n\,$ is uniformly bounded in $\,L^2$. Hence, upon taking a subsequence,
we may assume that there exists $\,u\in W^{1,2}(M,G)\,$ such that 
$\,u_n\rightharpoonup u\,$ in $\,W^{1,2}(M,G)$, $\,u_n\to u\,$ in $\,L^2(M,G)$. 
Note that this implies $\,a\,=\,u^{-1} b u\,+\,u^{-1}du\,$ and 
$\,{\underline u}_n\to {\underline u}\,$ in $\,L^2$ as well. 
From the definition of the  
invariant $\,c^k$, it is clear that distributional convergence of 
$\,\hbox{Tr}({\underline u}_n^{-1}d{\underline u}_n
\wedge {\underline u}_n^{-1}d{\underline u}_n
\wedge {\underline u}_n^{-1}d{\underline u}_n)\,$ 
to $\,\hbox{Tr}({\underline u}^{-1}d{\underline u}
\wedge {\underline u}^{-1}d{\underline u}
\wedge {\underline u}^{-1}d{\underline u})\,$ 
implies the convergence of the invariants. 
As in the proof of Lemma \ref{3inv} we have 

%%%%%%%%%%%%%%%%%%

$$
\begin{array}{c} 
\hbox{Tr}(u_n^{-1}du_n\wedge u_n^{-1}du_n\wedge u_n^{-1}du_n) = 
\hbox{Tr}(a_n\wedge a_n\wedge a_n - 3 a_n\wedge a_n\wedge u_n^{-1}b u_n \\
+ 3 u_n^{-1}b \wedge b u_n\wedge a_n  
- b \wedge b \wedge b)\,.
\end{array}
$$
Note that $\,da_n\,=\,-a_n\wedge a_n\,-\,F_B\,$ is bounded in $\,L^2$. 
The Bianci identity implies that $\,d(a_n\wedge a_n)=0$. By hypotheses, 
$\,a_n\rightharpoonup a\,$ and 
$\,a_n\wedge a_n\rightharpoonup a\wedge a\,$ in $\,L^2(M)$. Hence, 
$\,a_n\wedge a_n\wedge a_n\,$ converges to $\,a\wedge a\wedge a\,$ in the sense 
of distributions by the {\it div - curl\/} lemma. 
Since $\,u_n^{-1}\,$ converges strongly to $\,u^{-1}\,$ in $\,L^2\,$ 
and (taking a subsequence) 
$\,b u_n\,$ converges weakly to $\,bu\,$ in $\,L^2$, the product 
$\,u_n^{-1}b u_n\,$ converges to $\,u^{-1}b u\,$ in the sense of distributions.
However, $\,\|u_n^{-1}b u_n\|_{L^2}\,=\,\|u^{-1}b u\|_{L^2}\,=\,\|b\|_{L^2}$, 
so, upon taking a further subsequence, 
$\,u_n^{-1}b u_n\,$ converges to $\,u^{-1}b u\,$ weakly in $\,L^2$, 
and, hence, strongly in $\,L^2$. Since 
$\,a_n\wedge a_n\rightharpoonup a\wedge a\,$ in $\,L^2(M)$, 
we see that 
$\,\hbox{Tr}(a_n\wedge a_n\wedge u_n^{-1}b u_n)\,$ converges to 
$\,\hbox{Tr}(a\wedge a\wedge u^{-1}b u)\,$ distributionally. 
Similarly (on a further subsequence), $\,u_n^{-1}b \wedge b u_n\,$ 
converges to $\,u^{-1}b \wedge b u\,$ strongly in $\,L^2\,$ (recall: 
$\,b\wedge b\,\in L^2$). Since $\,a_n\rightharpoonup a$, we see that 
$\,\hbox{Tr}(u_n^{-1}b \wedge b u_n\wedge a_n)\,$ 
converges distributionally to $\,\hbox{Tr}(u^{-1}b \wedge b u\wedge a)$. 
Transition from $\,u\,$ to $\,{\underline u}\,$ is as in Lemma 
\ref{3inv}. 
This proves the lemma.

%%%%%%%%%%%%%%%%%%%%

We are now ready for the existence theorems.

%%%%%%%%%%%%%%%%%

\begin{theorem}\label{potentialexistence} 
Each sector $\,{\cal S}^{B,b,G}_{\rho, \alpha; c^1,\dots, c^N}[M]\,$ 
contains a minimizer of the Skyrme energy.
\end{theorem}
\noindent{\bf Proof.\/} Let 
$\,a_n\in {\cal S}^{B,b,G}_{\rho, \alpha; c^1,\dots, c^N}[M]\,$ 
be a minimizing sequence. As in the proof of the previous lemma, 
$\,da_n\,=\,-a_n\wedge a_n\,-\,F_B\,$ is bounded in $\,L^2\,$ and  
$\,d(a_n\wedge a_n)=0$. 
After taking a subsequence, we may assume that 
$\,a_n\rightharpoonup a\,$ in $\,L^2(M)\,$ and 
$\,a_n\wedge a_n\,$ converges weakly to some $2$-forms in $\,L^2(M,\frak g)$. 
By the {\it div - curl\/} lemma, 
$\,a_n\wedge a_n\,$ converges to $\,a\wedge a\,$ in the sense of distributions 
and, hence, weakly in $\,L^2$. By Lemma \ref{rhoweakconv}, 
$\,\rho_{B+a}\,=\,\rho_{B+a_n}\,=\,\rho$. 
As in the proof of   
Lemma \ref{cconv}, taking a subsequence, we produce a sequence 
$\,u_n\in W^{1,2}(M,G)\,$ weakly converging to $\,u\in W^{1,2}(M,G)\,$ 
such that $\,a_n= u_n^{-1} b u_n + u_n^{-1} du_n$, and 
$\,u_n^{-1} du_n\rightharpoonup u^{-1} du\,$  in $\,L^(M,\frak g)$. 
By definition, 
$\,\alpha[a]=\rho_{\tilde\theta+\hbox{pr}^*\tilde{\cal D}u}$, where 
$\,u\,$ satisfies $\,a= u^{-1} b u + u^{-1} du$. Applying Lemma 
\ref{rhoweakconv} to $\,\tilde\theta+\hbox{pr}^*\tilde{\cal D}u_n$ 
we conclude that $\,\alpha=\alpha[a_n]\to\alpha[a]$, so that 
$\,\alpha[a]=\alpha$. By Lemma \ref{cconv}, $\,c^k[a]=c^k$. 
This completes the proof.

\medskip
 
%%%%%%%%%%%%%%

\begin{corollary} Each sector $\,{\cal S}^G_{\alpha; c^1,\dots, c^N}(M)\,$ 
of the Skyrme maps contains a minimizer.  
\end{corollary} 
\noindent{\bf Proof.\/} By Remark \ref{D}, the map $\,{\cal D}\,$ maps 
$\,{\cal S}^G_{\alpha; c^1,\dots, c^N}(M)\,$ into 
$\,{\cal S}^{\theta, 0, G}_{{\bold 1},\alpha; c^1,\dots, c^N}[M]$. 
It is surjective by Lemma \ref{gaugeflat}. Also, $\,{\cal D}\,$ 
preserves energy, i.e., $\,E(u)=E[{\cal D} u]$. Any map $\,u\,$ 
in the  inverse image of a minimizer 
in $\,{\cal S}^{\theta, 0, G}_{{\bold 1},\alpha; c^1,\dots, c^N}[M]\,$ 
minimizes $\,E(\cdot)$. End of proof.

%

%
% BibTeX users please use
% \bibliographystyle{}
% \bibliography{}
%
% Non-BibTeX users please use

\end{document}